\documentclass[journal]{IEEEtran}
\usepackage{lineno}
\usepackage{subfig}
\usepackage{float}
\usepackage{amsmath}
\usepackage{graphicx}
\usepackage{multirow}
\usepackage{amsmath}
\usepackage{booktabs}
\usepackage{makecell}
\setcellgapes{2pt}
\usepackage{setspace}
\usepackage{adjustbox}
\usepackage{csquotes}
\usepackage[normalem]{ulem}
\usepackage{textcomp}
\usepackage{xcolor}
\usepackage{threeparttable}
\usepackage{amsmath,amssymb,amsfonts}
\usepackage{graphicx}
\usepackage{textcomp}

\usepackage{amstext}

\usepackage{array}

\usepackage{amssymb}
\floatstyle{ruled}

\usepackage{graphicx}
\usepackage{amsmath}
\usepackage{commath}
\usepackage{mathtools}
\usepackage{algorithm,algpseudocode}
\makeatletter
\def\hlinew#1{%
	\noalign{\ifnum0=`}\fi\hrule \@height #1 \futurelet
	\reserved@a\@xhline}

\definecolor{black}{rgb}{0,0,0}
\definecolor{red}{rgb}{1,0,0}
\definecolor{blue}{rgb}{0,0,1}

\modulolinenumbers[5]

\ifCLASSINFOpdf
\else
\fi
\begin{document}
	%
	\title{Blockchain-Federated-Learning and Deep Learning Models for COVID-19 detection using CT Imaging}
	%
	%
	%

	\author{Rajesh Kumar*,
		Abdullah Aman Khan,
		Sinmin Zhang, Jay Kumar, Ting Yang, Noorbakhsh Amiri Golilarz, Zakria, Ikram Ali, Sidra Shafiq and WenYong Wang
		\thanks{R. Kumar, A. A. Khan, W. Wang and Y. Ting are with the School of Computer Science and Engineering, University of Electronic Science and Technology of China, Chengdu, 611731, China.}
		\thanks{J. Kumar is with the Data Mining Lab, School of Computer Science and Engineering, University of Electronic Science and Technology of China, Chengdu, 611731, China}%
		\thanks{Corresponding author: WenYong Wang (wangwy@uestc.edu.cn) and Rajesh Kumar (rajakumarlohano@gmail.com)}}
	
	%
	%

	\markboth{}%
	{Shell \MakeLowercase{\textit{et al.}}: Bare Demo of IEEEtran.cls for IEEE Journals}
	%



	\maketitle
	
	\begin{abstract}
		With the increase of COVID-19 cases worldwide, an effective way is required to diagnose COVID-19 patients. The primary problem in diagnosing COVID-19 patients is the shortage and reliability of testing kits, due to the quick spread of the virus, medical practitioners are facing difficulty identifying the positive cases. The second real-world problem is to share the data among the hospitals globally while keeping in view the privacy concerns of the organizations. Building a collaborative model and preserving privacy are major concerns for training a global deep learning model. This paper proposes a framework that collects a small amount of data from different sources (various hospitals) and trains a global deep learning model using blockchain based federated learning. Blockchain technology authenticates the data and federated learning trains the model globally while preserving the privacy of the organization. First, we propose a data normalization technique that deals with the heterogeneity of data as the data is gathered from different hospitals having different kinds of CT scanners. Secondly, we use Capsule Network-based segmentation and classification to detect COVID-19 patients. Thirdly, we design a method that can collaboratively train a global model using blockchain technology with federated learning while preserving privacy. Additionally, we collected real-life COVID-19 patients’ data, which is, open to the research community. The proposed framework can utilize up-to-date data which improves the recognition of computed tomography (CT) images. Finally, our results demonstrate a better performance to detect COVID-19 patients.
	\end{abstract}
	

	%
	\IEEEpeerreviewmaketitle

\section{INTRODUCTION}

\subsection{Background}

A new type of Coronavirus emerged in the city of Wuhan in China. Unfortunately, within weeks this coronavirus ( COVID-19) speared to several countries and it has been proven fatal. With an estimated 325,000 deaths in 4 months, the COVID-19 virus is considered one of the most deadly viruses \cite{Coronavirus2020}. The first confirmed death from COVID-19 infection was recorded in early January this year. The Coronavirus family is categorized into 7 categories i.e. Human Coronavirus 229e (HCOV -229e), Human Coronavirus OC43 (hcov-OC43), SARS-CoV, Human Coronavirus NL63 (HcoV-NL63, New Haven Coronavirus), Human Coronavirus HKU1, Middle East Respiratory Syndrome Coronavirus (MERS-Cov), and Wuhan Coronavirus. This novel coronavirus is the seventh type ( COVID-19). Some coronavirus has mild symptoms while others such as SARS (severe acute respiratory or syndrome-related Coronavirus), and MERS (middle east respiratory) are much more dangerous. Coronavirus can be easily transmitted between humans mainly through social interaction with an active patient or direct contact with an infected animal.

Without any warning, the number of COVID-19 patients suddenly started to increase leaving the governments and medical practitioners unprepared to handle such a situation. Consequently, there is a shortage of testing kit supplies, and many hospitals worldwide are facing a challenge in identifying COVID-19 positive patients. The following criteria are used to diagnose COVID-19 patients: Clinical symptoms, Epidemiological history, and Positive CT and Pathogenic Testing. Radiological imaging is also one of the COVID-19s’ major diagnosis method. Most COVID-19 cases exhibit common features (visual symptoms) on CT images, including early ground-glass opacity, and late-stage pulmonary consolidation. There is also a rounded morphology and a peripheral lung distribution \cite{Chung2020,Huang2020}. While typical CT images may help to screen suspected COVID-19 cases at an early stage, CT images of various viral pneumonia are similar and overlap with other infectious and inflammatory lung diseases \cite{Choe2019,Kermany2018,Negassi2020,Gulshan2016}. It is worth noting that radiologists distinguish between COVID-19 and other viral pneumonia. The previous work focus on to diagnose the COVID-19 patients using computed tomography \cite{Wang2020,Chen2020,Li2020,Zheng2020,CovidGAN}. Therefore, these previous work do not focus on collaboratively learn model  and also do not consider the privacy issue of the hospitals.
\subsection{Motivations}
The motivation of our study is inspired by some fundamental problems. COVID-19 is spreading rapidly having different symptoms with different symptoms with different patients. Thus, hospitals can share their data for the accurate diagnosis of COVID-19 patients. Sharing data securely (without leakage the privacy of users)  and train the global  model for detection of the positive cases, is a challenging task. Moreover, the existing studies are not capable enough to share the data collaboratively and train the model accurately. Collecting data from various sources is a big challenge and a bottleneck in the advancement of AI-based techniques. The availability of such confidential data is not possible due to the absence of privacy-preserving approach for the health care centers \cite{9199262, 9203904, chiu2020semi, Lu2020, pang2020realizing, sun2020towards, yang2020federated, zhou2020cefl, xu2019verifynet, xu2018enabling}. Furthermore, to train the deep learning model collaboratively, over a public network, is another challenge.

The latest report of the World Health Organization reveals that COVID-19 is an infectious disease that primarily affects the lungs such as SARS, giving them a honeycomb-like appearance \cite{Zhang2020a}. Even after recovering from COVID-19, some patients have to live with permanent lung damage \cite{Ai2020}. First motivation of our work to find small infected areas in the lungs by COVID-19,  it benefits the professional radiologists do not missed infection. Second motivations to share the data to train a better deep learning model, while keeping in view the privacy concern of the data providers.  The advantage to share the data is feasible to develop a deep learning-based model for automatic detection of COVID-19.

\begin{itemize}
	\item The First challenge is the availability of  confidential data is not possible due to the absence of privacy. 
	\item  The second challenge is to train the global model (Federated model) via blockchain network.
	\item The third challenge is the unavailability of a dataset, it is quite challenging to collect enough amount of training data and make it better predication model with the privacy concerns of hospitals.
	\item Finally, to recognize the patterns of the lung screening of COVID-19 is also a challenging task.
\end{itemize}

\subsection{Our approach}

In this paper, we propose a framework that builds an accurate collabroative model using data from multipile hospitals to recognize CT scans of COVID-19 patients. The proposed blockchain based federated learning framework learns collaboratively from multiple hospitals having different kinds of CT scanners. Firstly, we propose a data normalization process to normalize the data obtained from the different sources. Then we employ deep learning models to recognize the COVID-19 patterns of lung CT scans. We use segcaps for image segmentation and further train  a Capsule Network \cite{DBLP:conf/nips/SabourFH17} for better generalization. We found the capsule network achieved better performance as compared to other learning models. Finally, we train the global model and solve the privacy issue using the federated learning technique. The proposed framework  collects the data and collaboratively trains an intelligent model then shares this intelligent model in a decentralized manner over the public network.

By using federated learning, the hospitals keep can their data private and share only weights and gradients while blockchain technology is used to distribute the data among the hospitals.
The decentralized architecture for data sharing among multiple hospitals shares the data securely without leakage the privacy of the hospitals.
Additionally, this article introduces a new dataset, named CC-19, related to the latest family of coronavirus i.e. COVID-19. The dataset contains the Computed Tomography scan (CT) slices for 89 subjects. Out of these 89 subjects, 68 were confirmed patients (positive cases) of the COVID-19 virus, and the rest 21 were found to be negative cases. The dataset contains 34,006 CT scan slices (images) belonging to 89 subjects. The data for these patients were collected on various days having about 231 CT scan volumes in total.

\subsection{Contributions}

The main contributions of the paper are not limited to:
\begin{enumerate}
	\item This paper proposes a data normalization technique (to accurately train the federated learning model) as the data is collected from different sources (i.e, Hospitals) and devices (CT scanner machines). 
	\item The proposed technique detects the patterns of COVID-19 from the lung CT scans using Capsule Network based segmentation and classification.
	\item This paper proposed a blockchain empowered method to collect the dataset collaboratively from different sources while keeping in view the organizations’ privacy concerns. Federated learning employed is to protect the organizations’ data privacy and train the global deep learning model using less accurate local models.
	
	\item Additionally, we introduce a new dataset that consists of 89 subjects out of which 68 subjects are confirmed COVID-19 patients. The dataset contains 34,006 CT scan slices (images) belonging to 89 subjects.

\end{enumerate}

\subsection{Applications}
The proposed approach is practical for big data analysis (i.e., lung CT scans), and it  efficiently process the data using  blockchain and deep learning model. Consider a scenario of the real-time use case of a hospital having some new symptoms of the COVID-19 virus. To find out new symptoms or new information regarding COVID-19, the data needs to be stored on a decentralized network without leakage of the privacy of the patients and securely share the knowledge of the latest symptoms.
The federated learning secures data through the decentralized network and distributes the training task to train a better model using the latest available patients data.

The proposed framework collects a small amount of data from various sources and to train the deep learning model, The blockchain combined the each trained model via federated learning. The trained model blockchain network provides more better and accurate  predication beacuse it holds the the newest information about COVID-19 symptoms.

\subsection{Structure of paper}
The rest of this paper is organized as follows: In Section II, this paper proposes a Capsule Network based segmentation and classification model and blockchain based federated learning for secure data sharing without leakage the privacy. In Section III, we describe the dataset and experiment results for our proposed scheme. In Section IV, we present an overview of the studies related to deep learning, COVID-19, and federated learning. Finally, Section V concludes this paper.

\section{Proposed Model \label{sec:Proposed-Model}}
In reality, hospitals and other relevant organizations are reluctant to share their patients’ data to preserve privacy of the patients. Moreover, it is a known fact that deep learning models required a large amount of data to train a model that can handle real-world problems. For that reason, this paper considers collecting multiple hospitals' data without leakage of data privacy. This paper proposes blockchain based federated learning framework to train and share a collaborative model. Federated learning is used to combine the weights of the locally trained model by the hospital referred as to a global or collaborative model.

As the data is collected from multiple sources, for that reason, we design a normalization technique to deal with different kinds of CT scanners (Brilliance ICT, Samatom definition Edge, Brilliance 16P CT) data. After normalization of the data, we segmented the images and then train the model for recognization of COVID-19 suspects using the Capsule Network.

We divided the methodology into two parts i) Local model ii) Federated learning. First, we solve the problem of heterogeneous CT scan data. Then, we use the Segcps \cite{lalonde2018capsules} for segmentation and train the local model to detect the patterns of COVID-19. Finally, we share the local model weights to the blockchain network to train the global model.

\subsection{Data Normalization}
A major issue with federated learning is to deal with input data from multiple sources and various machines with different parameters. Most of the existing techniques are not efficient enough to deal with this problem for federated learning.
To solve this issue, we propose a normalization technique that can deal with any CT scan and bring the images to the same standard. Because of this normalization, federated learning can deal with the heterogeneity of the dataset and train a better learning model. The normalization method has two phases i) spatial normalization, and ii) signal normalization. Spatial normalization deal with the dimension and resolution of the CT scan. Signal normalization deals with the intensity of each voxel of the CT scanners which is based on the lung window.

\subsubsection{Spatial Normalization}As already discussed, different CT scanners have different parameters for CT scans such as high-resolution scan volume is $0.31 \times 0.31 \times 0.31 \mathrm{~mm}^{3}$ and low resolution $0.98 \times 0.98 \times 2.5 \mathrm{~mm}^{3}$ . In our case, we used federated learning for the data obtained from multiple sources. We use the standardized volume $334 \times 334 \times 512 \mathrm{~mm}^{3}$ for human lung. Moreover, we use the Lanczos interpolation [27] to resale the standard resolutions.

\subsubsection{Signal Normalization} As every CT scan has Hounsfield Units (HU) and the data collected from different hospitals have different HU (i.e.,-400 HU to -600 HU). In medical practice, radiologists set the lung window for every CT scanner. There are different types of windows the one is window Level ($ WL $) and the other is window width ($ WW $) are mostly used. 
Where $ WL $ is defined as the central signal value and $ WW $ defines the width of this window. The proposed Equation \ref{eq:001} represents the upper bound and the lower bound of the voxel.

\begin{equation}
	\label{eq:001}
	I_{\text {normalized}}=\frac{I_{\text {original}}-\mathrm{WL}}{\mathrm{WW}}
\end{equation}

$I_{\text {original}}$ is the intensity of the data and $I_{\text {normalized}}$ is the final intensity. We set the range of the lung window is $[-0.5,0.5]$ to standardized the embedding space.

\subsection{Segmentation and Classification Model }
This section proposes the segmentation based on \cite{lalonde2018capsules}. Further, the Capsule Network is trained for the detection of COVID-19 using the segmented CT scan images. 

\subsubsection{Segmentation}
We take 2D slices for the segmentation. A  standardized volume $334 \times 334 \times 512 \mathrm{~mm}^{3}$ for human lung segmentation is used. Each CT scan volume (3D) has three planes $XY$, $XZ$, and $YZ$. We formalize the $XZ$ or $YX$ planes to easily differentiated the lung infection (as shown in first row of Figure \ref{fig:seg}).

\begin{equation}
	\label{Eq:above}
	\mathrm{prob}^{\mathcal{B}}=g\left(Pr_{\mathrm{xy}}^{\mathcal{B}}, {Pr}_{\mathrm{yz}}^{\mathcal{B}}, Pr_{\mathrm{xz}}^{\mathcal{B}}\right)
\end{equation}

Where ${prob}^{\mathcal{B}}$ defines as probability and ${\mathcal{B}}$ is the infection point. $g$ is the method to define the voxel of three dimensions views. $g$ is aggregation function to predict the $P_{xy}$, $P_{yz}$, and $P_{xz}$ voxel. Thus, the traditional Equation is time-consuming, so, we modify the Equation \ref{Eq:above} to:

\begin{equation}
	\begin{aligned}
		\hat{{prob}}^{\mathcal{B}} &=g\left(\hat{{pr}}_{{xy}}^{\mathcal{B}}, \hat{{pr}}_{{yz}}^{\mathcal{B}}, \hat{{pr}}_{{xz}}^{\mathcal{B}}\right) \\
		&=g\left( {f_{xy}^{\cal B}\left( {pr_{xy}^{\cal B}} \right),f_{yz}^{\cal B}\left( {pr_{yz}^{\cal B}} \right),f_{xz}^{\cal B}\left( {pr_{xz}^{\cal B}} \right)} \right)
	\end{aligned}
\end{equation}

\subsubsection{Capsule Networks for Classification of COVID-19}

%
A deep learning framework usually has a feature extraction pipeline that estimates and extracts prominent features. Afterward, a learning process such as MLP (multi-layer perceptron) is applied to learn the appropriate class on the extracted features. Over the past few years, researchers have used and fine-tuned the feature extraction pipeline of these robust deep learning frameworks. We design a Capsule Network because it achieves high performance in detecting diseases in the medical images. The previous technique needs lots of data to train a more accurate model. The Capsule Network improves the deep learning models' performance inside the internal layers of the deep learning models. The architecture of our modified Capsule Network is shown in Figure \ref{fig:capsuleNetwork}, which is similar to Hinton’s Capsule Network. The Capsule Network contains four layers: i)convolutional layer, ii) hidden layer, iii) PrimaryCaps layer, and iv) DigitCaps layer.

A capsule is created when input features are in the lower layer. Each layer of the Capsule Network contains many capsules. To train the Capsule Network, the activation layer represents instantiate parameters of the entity and compute the length of the Capsule Network to re-compute the scores for the feature part. Capsule Networks is a better replacement for Artificial Neural Network (ANN). Here, the capsule acts as a neuron. Unlike ANN where a neuron outputs a scalar value, Capsule Networks tend to describe an image at a component level and associate a vector with each component. The probability of the existence of a component is represented by this vector’s length and replaces max-pooling with "routing by agreement". As capsules are independents the probability of correct classification increases when multiple capsules agree on the same parameters. Every component can be represented by a pose vector $U_i$ rotated and translated by a weighted matrix $W_{i,j}$ to a vector $\hat{u}_{i | j}$.
Moreover, the prediction vector can be calculated as:

\begin{figure*}[!tp]
	\centering
	\includegraphics[width=0.9\textwidth]{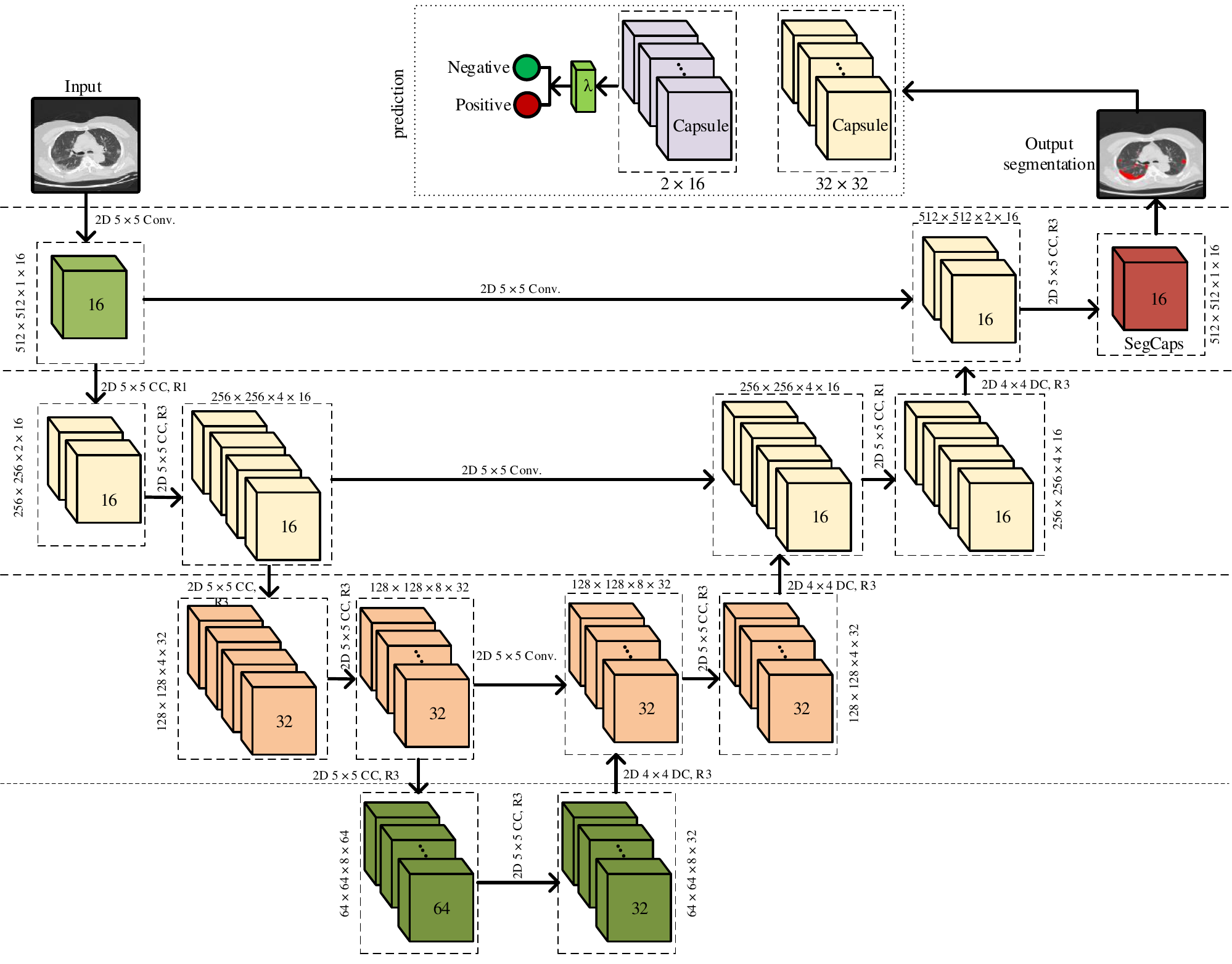}
	\caption{Our proposed, capsule-based (SegCaps) segmentation and classification model. Further, CC, DC, and Conv. represent convolution capsule, Deconvolution capsule, and convolution respectively.}
	\label{fig:capsuleNetwork}
\end{figure*}

\begin{equation}
	\hat{u}_{i|j}=W_{i,j}u_{i}
\end{equation}
The next higher level capsule i.e. $s_{j}$ processes the sum of predictions
from all the lower level capsules with $c_{i,j}$ as a coupling coefficient.
Capsules $s_{j}$ can be represented as:
\begin{equation}
	S_{j}=\sum_{i}c_{i,j}\hat{u}_{i|j}
\end{equation}
where $c_{i,j}$ can be represented as a routing softmax function
given as:
\begin{equation}
	c_{i,j}=\frac{e^{b_{ij}}}{\sum_{k}e^{b_{ik}}}
\end{equation}
As can be seen from the Figure \ref{fig:capsuleNetwork}, the parameter
c, A squashing function is applied to scale the output probabilities
between 0 and 1 which can be represented as:
\begin{equation}
	a=\frac{\|a\|^{2}}{1+\|a\|^{2}}\frac{a}{\|a\|}
\end{equation}
For further details, refer to the original study \cite{DBLP:conf/nips/SabourFH17}.
We perform the routing by agreement using the ~Algorithm \ref{alg:routing}

\begin{algorithm}[h]
	
	\begin{algorithmic}[1]
		
		\scriptsize
		\State Forall capsules $ i $ in layer $ l $ and capsule in layer $ l +1 $) do $ b_{i,j} \xleftarrow{} 0 $
		\State For k iterations do
		\State ~Forall capsule $ i $ in layer $ l $ do $ c_{i,j} $
		\State ~Forall capsule $ j $ in layer $ l + 1 $ do $ S_j $
		\State ~Forall capsule $ j $ in layer $ l + 1 $ do $ S_j $
		State ~Forall capsule $ i $ in layer $ l, j $ in layer $ l + 1 $ do $b_{i,j}\xleftarrow{} b_{i,j}+\hat{u}_{i|j}.v_j $
		\State Return $ v_j $
		
	\end{algorithmic}
	
	\centering{}\caption{Routing algorithm. \label{alg:routing}}
\end{algorithm}

C is an array after softmax, and it can be determined by dynamic routing by agreement. There are quite a few introductions to this method, the main meaning is that through several iterations, the distribution of the output of the low-level capsule to the high-level capsule is gradually adjusted according to the output of the high-level capsule, and finally an ideal distribution will be reached. The detailed training algorithm is shown in the paper \cite{Wang}. We use the Capsule Network to train the model and compare it with the state of art deep learning networks. Table 1 shows the difference between traditional and Capsule Network. In section \ref{experiments}, we compare traditional deep learning with the Capsule Network classifiers.

\begin{table}[]
	\label{fig:capsule-vs-traditional}
	\caption{A comparison between capsule and traditional neural network.}
	\begin{tabular}{>{\raggedright}p{0.2\columnwidth}lll}
		\toprule 
		\hline
		\rule[-1ex]{0pt}{3.5ex}Operation & Neuron (scalar) & Capsule (vector) \\ \hline
		\rule[-1ex]{0pt}{3.5ex}Affine transformation & NA&$\hat{u}_{i|j}=W_{i,j}u_{i}$ \\ \hline
		\rule[-1ex]{0pt}{3.5ex}Weighted sum &$a_{j}=\sum_{i=1}^{3} W_{i} x_{i}+b $& $ S_{j}=\sum_{i}c_{i,j}\hat{u}_{i|j} $\\ \hline
		\rule[-1ex]{0pt}{3.5ex}Activation & $ h_{w,b}(x)=f(a_{j} $)& $a_{j}=\frac{\|a_{j}\|^{2}}{1+\|a_{j}\|^{2}}\frac{a_{j}}{\|a_{j}\|} $ \\ \hline
		\rule[-1ex]{0pt}{3.5ex}Output & scalar & vector($ a_{j} $) \\ \hline
		\rule[-1ex]{0pt}{3.5ex}Graphical representation&{\raisebox{-\totalheight}{\includegraphics[width=0.3\columnwidth, height=10mm]{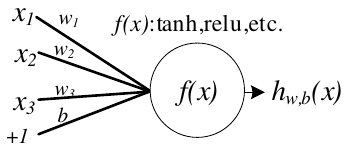}}} &\raisebox{-\totalheight}{\includegraphics[width=0.3\columnwidth, height=10mm]{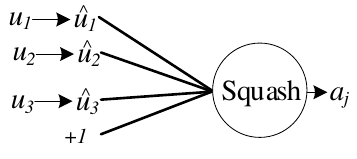}}\\ \hline
		\bottomrule	
	\end{tabular}
\end{table}

\subsection{Federated Learning to train the global model}

\begin{figure}[!tp]
	\centering
	\includegraphics[width=0.99\columnwidth]{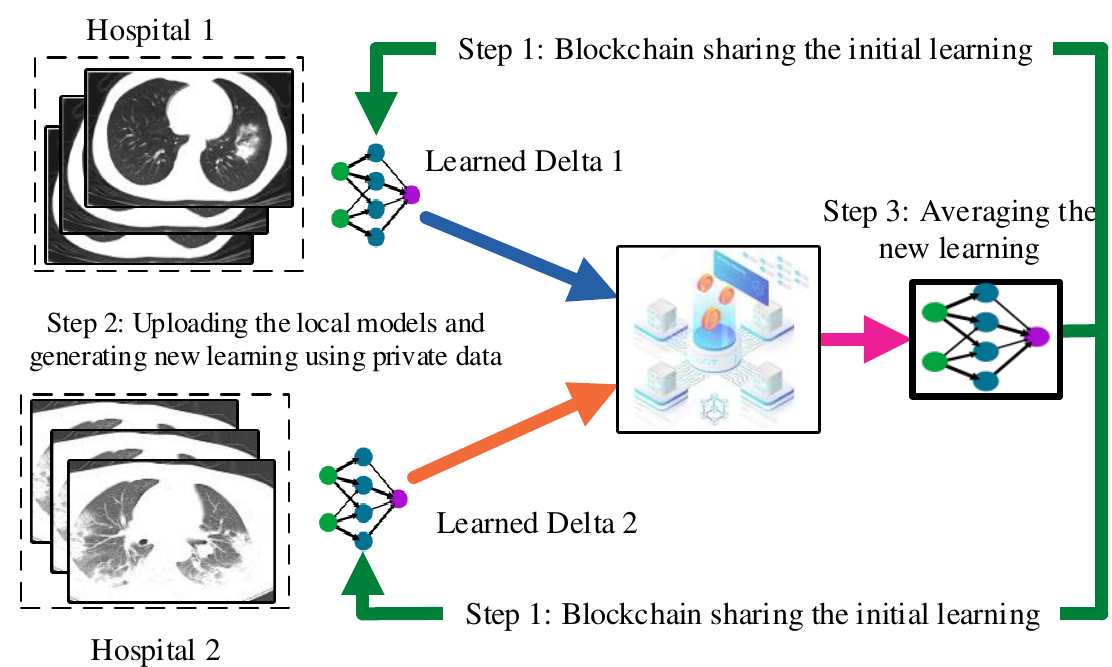}
	\caption{Overview of the Federated Learning process.}
	\label{fig:Federated-Learning}
\end{figure}

In this section, we consider a decentralized data sharing scenario with multiple hospitals. Each hospital is willing to share its model (weights), our proposed method assists in hiding the user data and share the model over a decentralized network. Further, federated learning is used to combine the net effect of different models shared by different hospitals. The base architecture of federated learning is shown in Figure \ref{fig:Federated-Learning}. The main goal is to utilize federated learning to share the data among the hospitals without leakage of privacy. We consider $H$ is the hospitals and $d$ is the union dataset. Each of the hospital $H$ agrees to share the data without leakage of private information. First, we train the global model $M$, without leaking the privacy, then we alter a small part of the randomized mechanism through 1) Random sub-sampling, 2) Distorting. The random sub-sampling model can get final weights $R(M)$ and share the data globally.
1) Random sub-sampling: Let $H$ be the number of hospitals. In every
round of the communication, a subset of $X_{t}$ of size $m_{t}\leq H$
is sampled. Then, distribute the weights ($wt$) among the hospitals.
The blockchain stores the local hospitals models $\left\{ w^{H}\right\} _{H=0}^{m_{t}}$
. The difference between the local and distributed model is referred
to as $H's$ update $\Delta w^{k}=w^{k}-w_{t}$. The updated
weights are sent to the decentralized network for each round.

2) Distorting: A Gaussian method was utilized to disorder the sum of updates. It requires information about the sensitivity information to sum all operations. The sensitivity of the updated version is measured by $\Delta w^{h}=\Delta w^{h}/\max\left(1,\frac{\left\Vert \Delta w^{h}\right\Vert _{2}}{S}\right)$. Scaling helps to ensure the limited second standard $\forall H,\left\Vert \Delta\bar{w}^{H}\right\Vert _{2}<S$. The sensitivity of the update bound operation by $S$. The updated model is defined as:

\begin{equation}
	\begin{array}{l}
		{w_{t+1}}={w_{t}}+\frac{1}{{m_{t}}}\underbrace{\sum\limits _{H=0}^{{m_{t}}}\Delta{w^{H}}/\max\left({1,\frac{{{\left\Vert {\Delta{w^{H}}}\right\Vert }_{2}}}{S}}\right)}_{{\rm {Gaussian \; mechanism\; approximation \; sum\;of\;updates}}}\\
		+\underbrace{{\cal N}\left({0,{\sigma^{2}}{S^{2}}}\right)}_{{\rm {Sum\;of\;update\;sclipped\;at\;}}S}
	\end{array}\label{eq:7}
\end{equation}

We noticed that the distortion of $1/mt$ in the Gaussian process is regulated by the $S^{2}\sigma^{2}/m$ noise variance. But this distortion should not surpass a certain amount. Otherwise, the additional noise removes too much detail from the sub-sampled average and no learning improvement can be gained. Gaussian mechanism and sub-sampling are distributed processes. Nevertheless, it is used for gradient averaging covering a single data point gradient at each iteration. This $m$ and $\sigma$ often describes the lack of privacy suffered when the randomized process produces an average estimate.

The preference will then be based on an upper limit at the r-distortion rate and a lower limit on the number of data provider sub-samples. Therefore describe the $Vc$ variance between patients as a measure of similarity among hospital updates shown in the below Equation. The parameters $(x,y)$ is the throughout of $H$ hospitals is defined as:

\begin{equation}
	VAR\left[\Delta w_{x,y}\right]=\frac{1}{H}\sum_{H=0}^{H}\left(\Delta w_{x,y}^{H}-\mu_{x,y}\right)^{2}\label{eq:8}
\end{equation}

where $\mu_{x,y}=\frac{1}{H}\sum_{H=1}^{H}\Delta w_{x,y}^{H}$

$V_{c}$ is defined as the sum of all variances in the update matrix:

\begin{equation}
	V_{c}=\frac{1}{b\times a}\sum_{x=0}^{q}\sum_{y=0}^{p}VAR\left[\Delta w_{x,y}\right]\label{eq:9}
\end{equation}

Finally, the $U_{s}$ update can be expressed as:

\begin{equation}
	U_{s}=\frac{1}{b\times a}\sum_{x=0}^{q}\sum_{y=0}^{p}\mu_{x,y}^{2}\label{eq:10}
\end{equation}

$\Delta w_{x,y}$ describes the $(x,y)-th$ parameters of the updates
from $\Delta w\in\mathbb{R}^{b\times a}$ for the communication round.
Moreover, $S$ defines trade-off. If $S$ has a smaller value then the noise
will be smaller.

\subsection{Blockchain based fast and effective Federated Learning}

\begin{figure}[!tp]
	\centering
	\includegraphics[width=0.99\columnwidth]{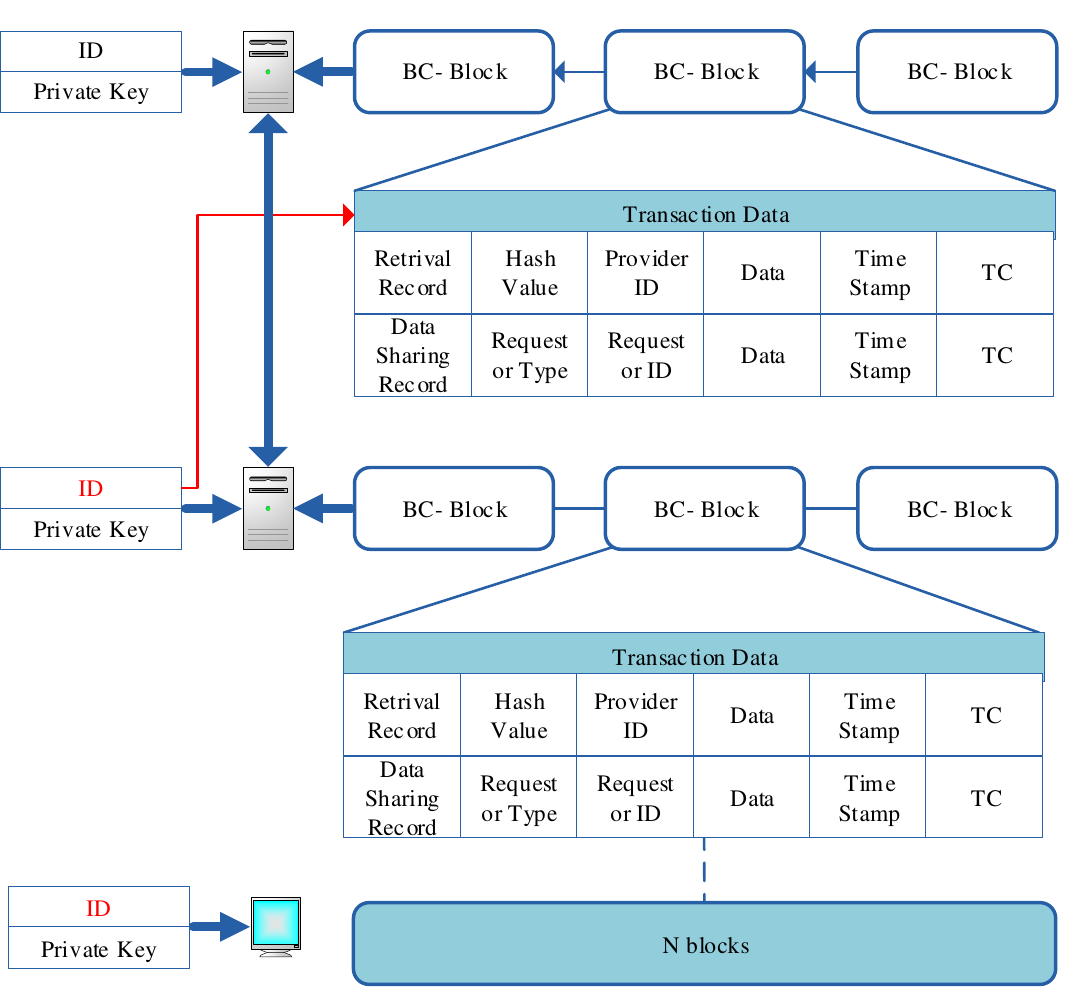}
	\caption{Overview of the Blockchain record storing process. TC represents the transaction count.}
	\label{fig:recordsOfBlockchain}
\end{figure}

As patients’ data is sensitive and the volume is high, placing data on the blockchain with its limited storage space is expensive both financially and for computational resources. Thus, the actual CT scan data is stored by the hospital, while, blockchain helps to retrieve the trained model. When a new hospital provides the data, it stores a transaction in the block to verify the owner of the data. The hospital data include the type of data and the size of the data. Each transaction for data sharing and retrieval process is shown in Figure \ref{fig:recordsOfBlockchain}. The proposed model solves data-sharing retrieval requests. Multiple hospitals can collaboratively share the data and train the collaborative model which can be used to predict optimal results. The retrieval mechanism does not violate the privacy of hospitals.
Inspired by Maymounkov et al. \cite{Maymounkov2002}, we present multi-organization architecture using blockchain technology. All hospitals $H$ are partitioned share data in various categories. Each category has a different community. Each community maintains the log table $Log(n)$. The blockchain stores the all unique IDs for every hospital.

Retrieving data into the physically present nodes is expressed by Equation \ref{eq:1}. We measure the distance between two nodes as Equation \ref{eq:1}, where, $H$ is the data categories to retrieve the data among the hospitals. Moreover, the distance of two nodes $d_{i}\left(H_{i},H_{j}\right)$ measured to the retrieve of data, and $\left(x_{pq}^{H_{i}}+x_{pq}^{H_{j}}\right)$ are the attributes of the weight matrix for the node $H_{i}$ and $H_{j}$, respectively. Every hospital generates its unique ID according to the logic and distance of the nodes.

\begin{equation}
	\begin{aligned}d_{i}\left(H_{i},H_{j}\right)= & \frac{\sum_{p,q\in\left\{ Hi\cup H_{j}-H_{i}\cap H_{j}\right\} }\left(x_{pq}^{H_{i}}+x_{pq}^{H_{j}}\right)}{\sum_{pq\in H_{i}\cup H_{j}}\left(x_{pq}^{H_{i}}+x_{pq}^{H_{j}}\right)}\\
		& \cdot\log\left(d_{p}\left(H_{i},H_{j}\right)\right)
	\end{aligned}
	\label{eq:1}
\end{equation}

Provided two nodes $H_{i}$ and $H_{j}$ with unique IDs $H_{i}(id)$
and $H_{j}(id)$ shown in the Equation \ref{eq:2}.

\begin{equation}
	d\left(H_{i},H_{j}\right)=H_{i}(\mathrm{id})\oplus H_{j}(\mathrm{id})\label{eq:2}
\end{equation}

To secure the privacy of data in a decentralized manner the randomized method for two hospitals nodes is shown in Equation \ref{eq:3}. Where $R$ and$R^{\prime}$ is the neighboring records of data. $O$ is the outcome set of data. $\mathcal{A}(R)\in S$ achieves the privacy of the data.

\begin{equation}
	Hr[\mathcal{A}(R)\in S]\leq\exp(\epsilon)\cdot Hr\left[\mathcal{A}\left(R^{\prime}\right)\in O\right]\label{eq:3}
\end{equation}

However, to achieve data privacy for multiple hospitals, Laplace is applied for the local model training $(m_{i})$:

\begin{equation}
	\hat{m}_{i}=m_{i}+\text{Laplace}(s/\epsilon)\label{eq:4}
\end{equation}

where $ s $ shows the sensitivity as expressed by Equation \ref{eq:5}:

\begin{equation}
	s=\max_{H,H^{\prime}}\left\Vert f(H)-f\left(H^{\prime}\right)\right\Vert _{1}\label{eq:5}
\end{equation}

The consensus algorithm is executed to train the global model by using the local models. As all nodes collaboratively train the model, we provide proof of work to share the data between the different nodes. During the training phase, the consensus algorithm checks the quality local models and the accuracy is measured by mean absolute error (MAE). $F(x_{i})$ shows predicated data and $m_{i}$, $y_{i}$ is the original data. The high accuracy of $m_{i}$ shows the low mean absolute error of $m_{i}$. The voting process consensus algorithm among the hospitals shown in Equation \ref{eq:6} and \ref{eq:11}. Where Equation \ref{eq:6} $MAE\left(m_{i}\right)$ shows the locally trained model and $\gamma$ shows the global models weights in Equation \ref{eq:11}.

\begin{equation}
	MAE\left(m_{i}\right)=\frac{1}{n}\sum_{i=1}^{n}\left|y_{i}-f\left(x_{i}\right)\right|\label{eq:6}
\end{equation}

\begin{equation}
	MAE\left(H_{j}\right)=\gamma\cdot MAE\left(m_{j}\right)+\frac{1}{n}\sum MAE\left(m_{i}\right)\label{eq:11}
\end{equation}

To preserve the hospitals' data privacy, all data is encrypted and signed using public and private keys $\left(\mathrm{PK}_{i},\mathrm{SK}_{i}\right)$, MAE calculates all transactions and broadcast $\left(H_{j}\right)$. $MAE(M)$ calculates each transaction of the model. If all transactions are approved then the record is stored in the distributed ledger. More precisely, the training of the consensus algorithm describes as follows:

\begin{enumerate}
	\item Node $H_{i}$ transfers the local model $m_{i}$ transaction to the
	$H_{j}$.
	\item Node $H_{j}$ transfers the local model $m_{i}$ to the leader.
	\item The leader broadcast the block the node to the $H_{i}$ and $H_{j}$.
	\item Verify the $H_{i}$ and $H_{j}$ and wait for the approval.
	\item Finally, store the blocks in the retrieval blockchain database.
\end{enumerate}

\subsubsection{Data Sharing Process}

\begin{figure}[!tp]
	\centering
	\includegraphics[width=0.99\columnwidth]{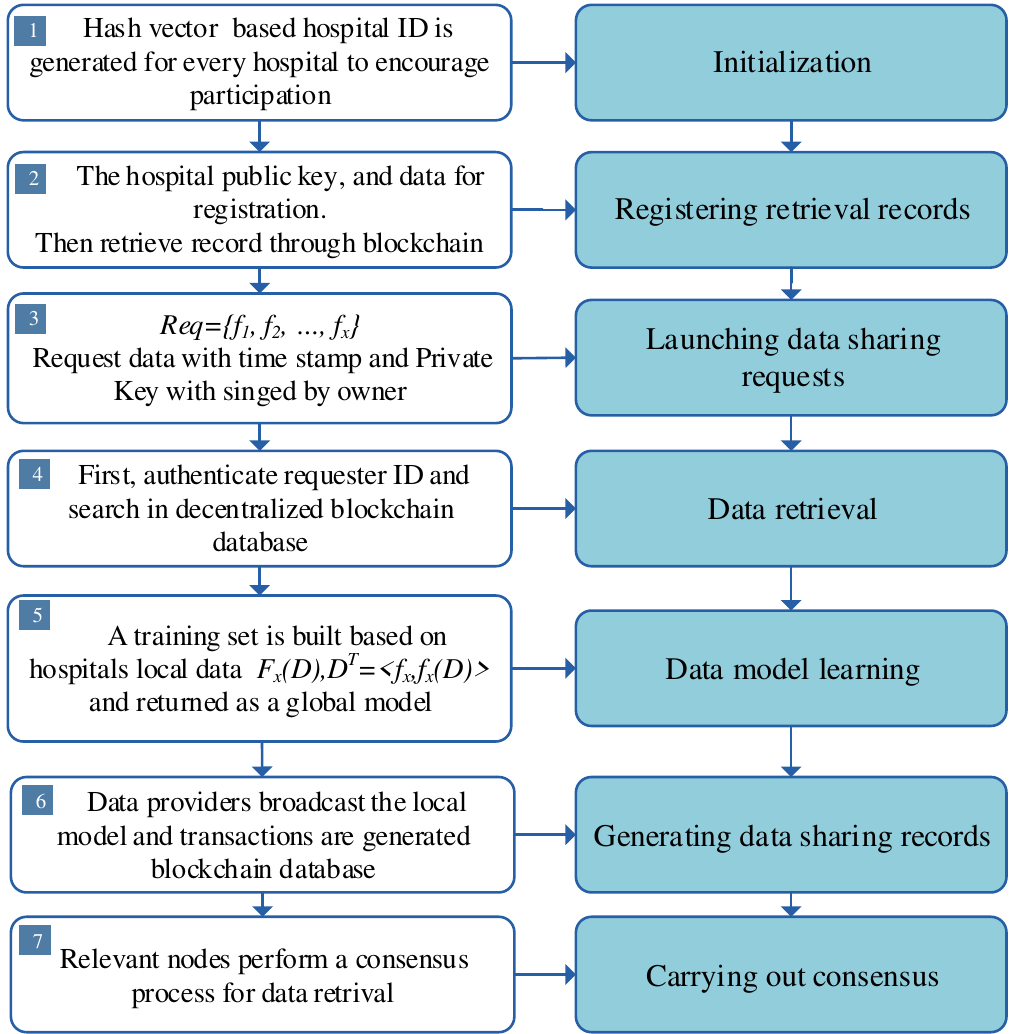}
	\caption{Data Sharing Process.}
	\label{fig:Data-Sharing-Process}
\end{figure}

Current approaches use encryption to protect data. It is a risk for data providers to share personal data because of certain security attacks. A simple solution is to transmit the data to the requester with legitimate details and to preserve the data holders’ privacy. Instead of sharing original data, data providers such as hospitals, exchange only the learned models with the requester. Figure \ref{fig:Data-Sharing-Process} shows the process of data sharing. The nodes are communicating with each other and the consensus process learns from federated data. The provider, and requester search and store the data into the blockchain nodes. More precisely, the steps of data sharing are shown in Figure \ref{fig:Data-Sharing-Process}. To integrate the blockchain with federated learning retrieved data securely for the multiple world-wide hospitals which can provide an effective prediction.

\begin{figure}[!tp]
	\centering
	\includegraphics[width=0.99\columnwidth]{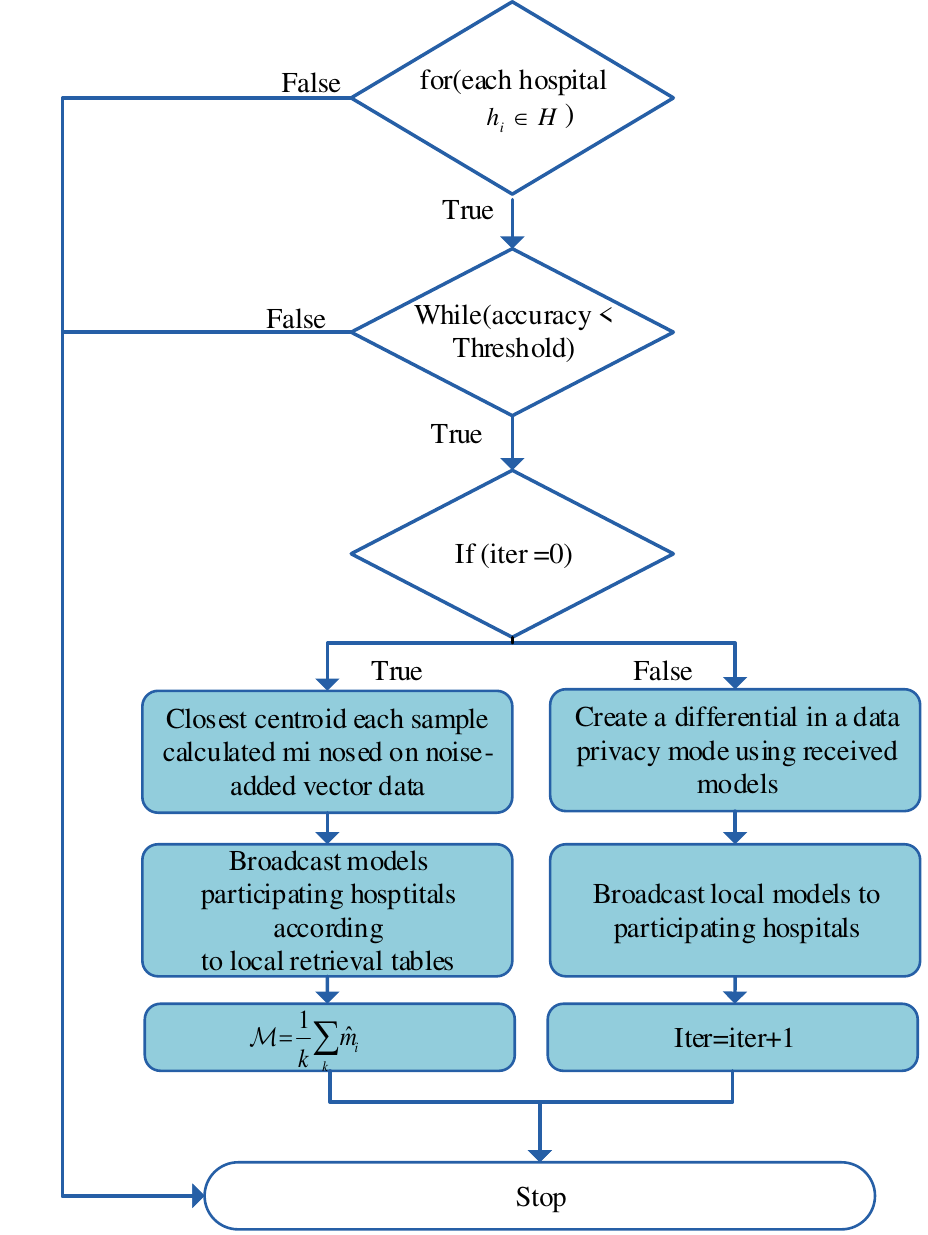}
	\caption{Private Federated Learning Algorithm.}
	\label{fig:Private-Federated-Learning}
\end{figure}

To protect the privacy of the data, we share the trained model instead of the original image data. The objective of the proposed architecture is to train the global model by using locally trained models. The secure data sharing is illustrated in Figure \ref{fig:Private-Federated-Learning}. In the first phase, we select the training data and then use the private federated learning algorithm for collaborative multi-hospital learning. In other words, the hospital shares the locally trained model weights to the blockchain network and federated learning combines the local model into a global model.

\section{Experiments Results}
\label{experiments}

\subsection{CC-19 Dataset}

\begin{figure}[!tp]
	\centering
	\includegraphics[width=0.99\columnwidth]{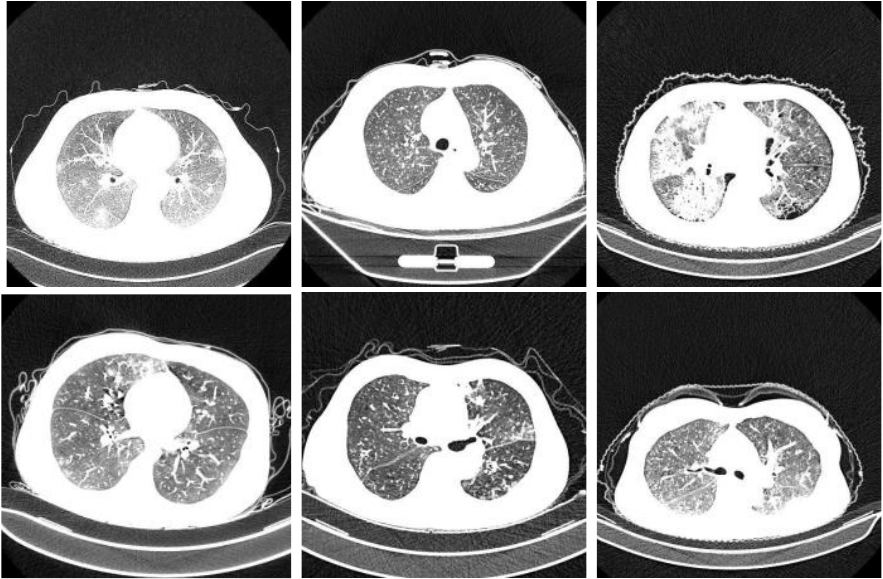}
	\caption{Some random samples of CT scan 2D slices taken from CC-19 dataset.}
	\label{fig:2dssampels}
\end{figure}

\begin{figure*}
	\centering
	\includegraphics[scale=0.7]{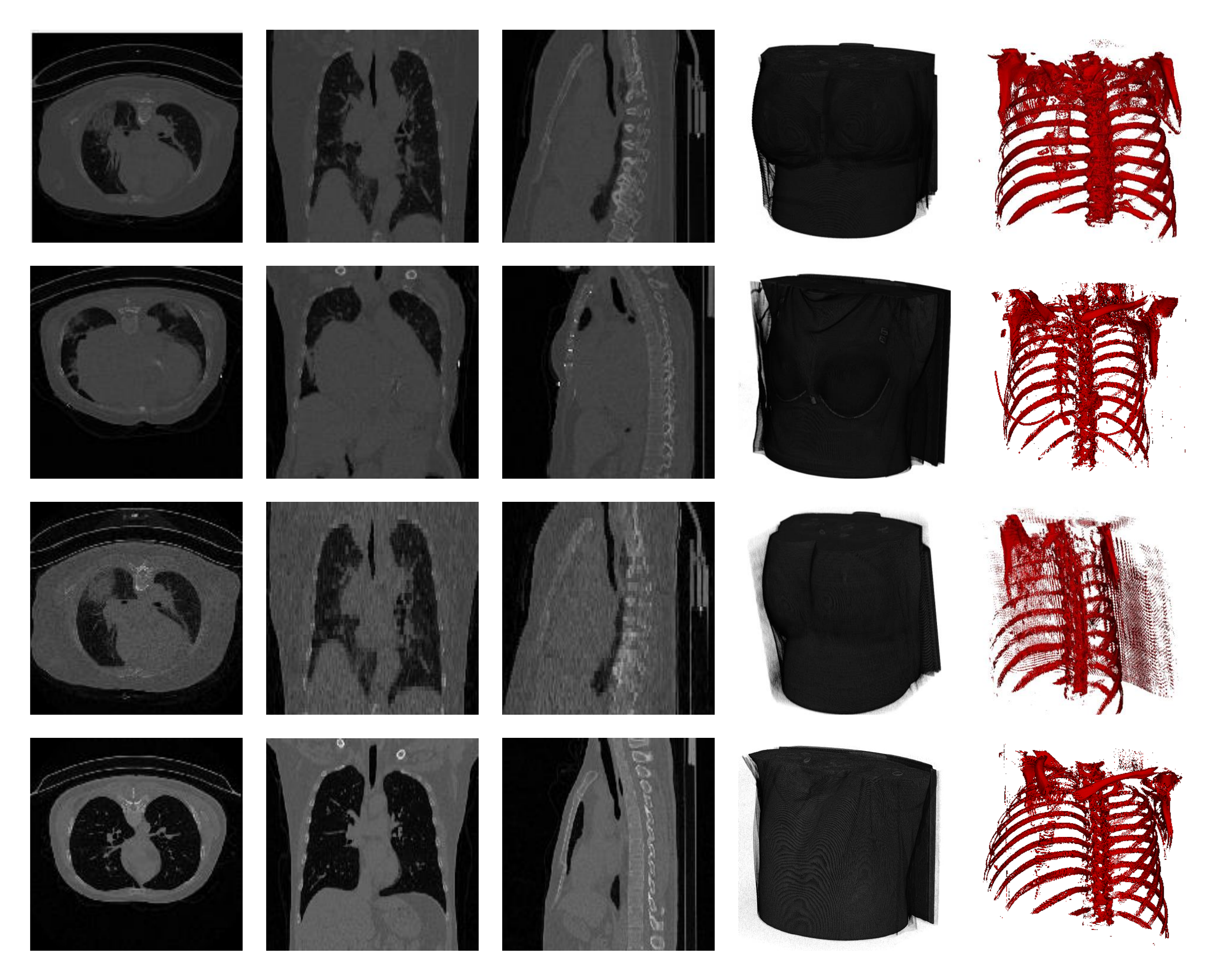} \caption{This figure shows some selected samples from the ``CC-19 dataset". Each row represents different patient samples with various
		Hounsfield Unit (HU) for CT scans. The first there columns represent the XY, XZ, and YX plane of the 3D- volume respectively. The fourth column represents a whole 3D-Volume followed by a bone structure in the fifth column.}
	\label{fig:3dssampels}
	
\end{figure*}

In the past, Artificial intelligence
(AI) has gained a reputable position in the field of clinical medicine.
And in such chaotic situations, AI can help the medical practitioners
to validate the disease detection process, hence increasing the reliability
of the diagnosis methods and save precious human lives. Currently,
the biggest challenge faced by AI-based methods is the availability
of relevant data. AI cannot progress without the availability of abundant
and relevant data.
In this paper, we collected the data CT scan data for 34006 slices from the 3 different hospitals. The data is scanned by 6 different scanners shown in Table \ref{tab:dataset}. In addition, we collected the third party dataset \cite{rahm2020,zhao2020COVID-CT-Dataset} from different sources to validate the federated learning methods. Moreover, the collected dataset is publicly available via GitHub (https://github.com/abdkhanstd/ COVID-19). The collected data set contains the Computed Tomography
scan (CT) slices for 89 subjects. Out of these 89 subjects, 68 were
confirmed patients (positive cases) of the COVID-19 virus, and the
rest 21 were found to be negative cases. The proposed dataset CC-19
contains 34,006 CT scan slices (images) belonging to 89 subjects out
of which 28,395 CT scan slices belong to positive COVID-19 patients.
Figure \ref{fig:2dssampels} shows some 2D slices taken from CT scans
of the CC-19 dataset. Moreover, some selected 3D samples from the dataset
are shown in Figure \ref{fig:3dssampels}. The Hounsfield unit (HU)
is the measurement of CT scans radiodensity as shown in Table \ref{tab:hu}.
Usually, CT scanning devices are carefully calibrated to measure the
HU units. This unit can be employed to extract the relevant information
in CT Scan slices. The CT scan slices have cylindrical scanning bounds.
For unknown reasons, the pixel information that lies outside this
cylindrical bound was automatically discarded by the CT scanner system.
But fortunately, this discarding of outer pixels eliminates some steps
for preprocessing.

\begin{table*}
	\caption{CC-19 dataset collected from three different hospitals (A, B, and C).}
	\label{tab:dataset}
	
	\begin{tabular}{l>{\raggedright}p{0.1\textwidth}>{\raggedright}p{0.1\textwidth}>{\raggedright}p{0.1\textwidth}>{\raggedright}p{0.1\textwidth}>{\raggedright}p{0.1\textwidth}>{\raggedright}p{0.1\textwidth}}
		\toprule
		\hline 
		\toprule
		Hospital ID & A & A & B & B & C & C\tabularnewline
		\hline 
		CT scanner ID & 1 & 2 & 3 & 4 & 5 & 6\tabularnewline
		\hline 
		Number of Patients & 30 & 10 & 13 & 7 & 20 & 9\tabularnewline
		\hline 
		Infecation annotation & Voxel-level & Voxel-level & Voxel-level & Voxel-level & Voxel-level & Voxel-level\tabularnewline
		\hline 
		CT scanner & SAMATOM scope & Samatom Definitation Edge & Brilliance 16P iCT & Brilliance iCT & Brilliance iCT & GE 16-slice CT scanner\tabularnewline
		\hline 
		Lung Window level (LW) & -600 & -600 & -600 & -600 & -600 & -500\tabularnewline
		\hline 
		Lung Window Witdh (WW) & 1200 & 1200 & 1600 & 1600 & 1600 & 1500\tabularnewline
		\hline 
		Slice thickness (mm) & 5 & 5 & 5 & 5 & 5 & 5\tabularnewline
		\hline 
		Slice increment (mm) & 5 & 5 & 5 & 5 & 5 & 5\tabularnewline
		\hline 
		Collimation(mm) & 128{*}0.6 & 16{*}1.2 & 128{*}0.625 & 16{*}1.5 & 128{*}0.6 & 16{*}1.25\tabularnewline
		\hline 
		Rotation time (second) & 1.2 & 1.0 & 0.938 & 1.5 & 1.0 & 1.75\tabularnewline
		\hline 
		Pitch & 1.0 & 1.0 & 1.2 & 0.938 & 1.75 & 1.0\tabularnewline
		\hline 
		Matrix & 512{*}512 & 512{*}512 & 512{*}512 & 512{*}512 & 512{*}512 & 512{*}512\tabularnewline
		\hline 
		Tube Voltage (K vp) & 120 & 120 & 120 & 110 & 120 & 120\tabularnewline
		
		\hline 
		\bottomrule
	\end{tabular}
\end{table*}

\begin{table}[h]
	\begin{tabular}{ccc}
		
		\toprule
		\hline
		S/No & Substance & Hounsfield Unit (HU)\tabularnewline
		\hline
		1 & Air & -1000\tabularnewline
		\hline
		2 & Bone & +700 to +3000\tabularnewline
		\hline
		3 & Lungs & -500\tabularnewline
		\hline
		4 & Water & 0\tabularnewline
		\hline
		5 & Kidney & 30\tabularnewline
		\hline
		6 & Blood & +30 to +45\tabularnewline
		\hline
		7 & Grey matter & +37 to +45\tabularnewline
		\hline
		8 & Liver & +40 to +60\tabularnewline
		\hline
		9 & White matter & +20 to +30\tabularnewline
		\hline
		10 & Muscle & +10 to +40\tabularnewline
		\hline
		11 & Soft Tissue & +100 to +300\tabularnewline
		\hline
		12 & Fat & -100 to -50\tabularnewline
		\hline
		13 & Cerebrospinal fluid(CSF) & 15\tabularnewline
		\hline
		\bottomrule
	\end{tabular}
	
	\caption{Various values of Hounsfield unit (HU) for different substances.\label{tab:hu}}
	
\end{table}

Collecting dataset is a challenging task as there are many ethical
and privacy concerns observed the hospitals and medical practitioners.
Keeping in view these norms, this dataset was collected in the earlier
days of the epidemic from various hospitals in Chengdu, the capital
city of Sichuan. Initially, the dataset was in an extremely raw form.
We preprocessed the data and found many discrepancies with most of
the collected CT scans. Finally, the CT scans, with discrepancies,
were discarded from the proposed dataset. All the CT scans are different
from each other i.e. CT scans have a different number of slices for
different patients. We believe that the possible reasons behind the
altering number of slices are the difference in height and body structure
of the patients. Moreover, upon inspecting various literature, we
found that the volume of the lungs of an adult female is, comparatively,
ten to twelve percent smaller than a male of the same height and age
\cite{DBLP:journals/AJRCCM/Bellemare}.

\subsection{Evaluation Measures}

Specificity and sensitivity are the abilities of a model that how correctly the model identifies a subject with disease and without a disease. In our case, it is critical to detect a COVID-19 patient as missing a COVID-19 patient can have disastrous consequences. The formulas of the measures are given as follows:

\[
\text{Precision}=\frac{TP}{TP+FP}
\]

\[
\text{sensitivity=recall}=\frac{TP}{TP+FN}
\]

\[
\text{specificity}=\frac{TN}{TN+FP}
\]

\[
\text{Total accuarcy}=\frac{TP + TN}{TP+TN+FP+FN}
\]
A medical diagnosis based system needs to have high sensitivity and recall. We present a comprehensive overview of various famous deep learning frameworks. The results presented in Table \ref{tab:True-Positive-local-model} indicate the superiority of our proposed method.

\subsection{Results of the pattern recognition with the benchmark algorithms}

\begin{figure}[!tp]
	\centering
	\includegraphics[width=0.99\columnwidth]{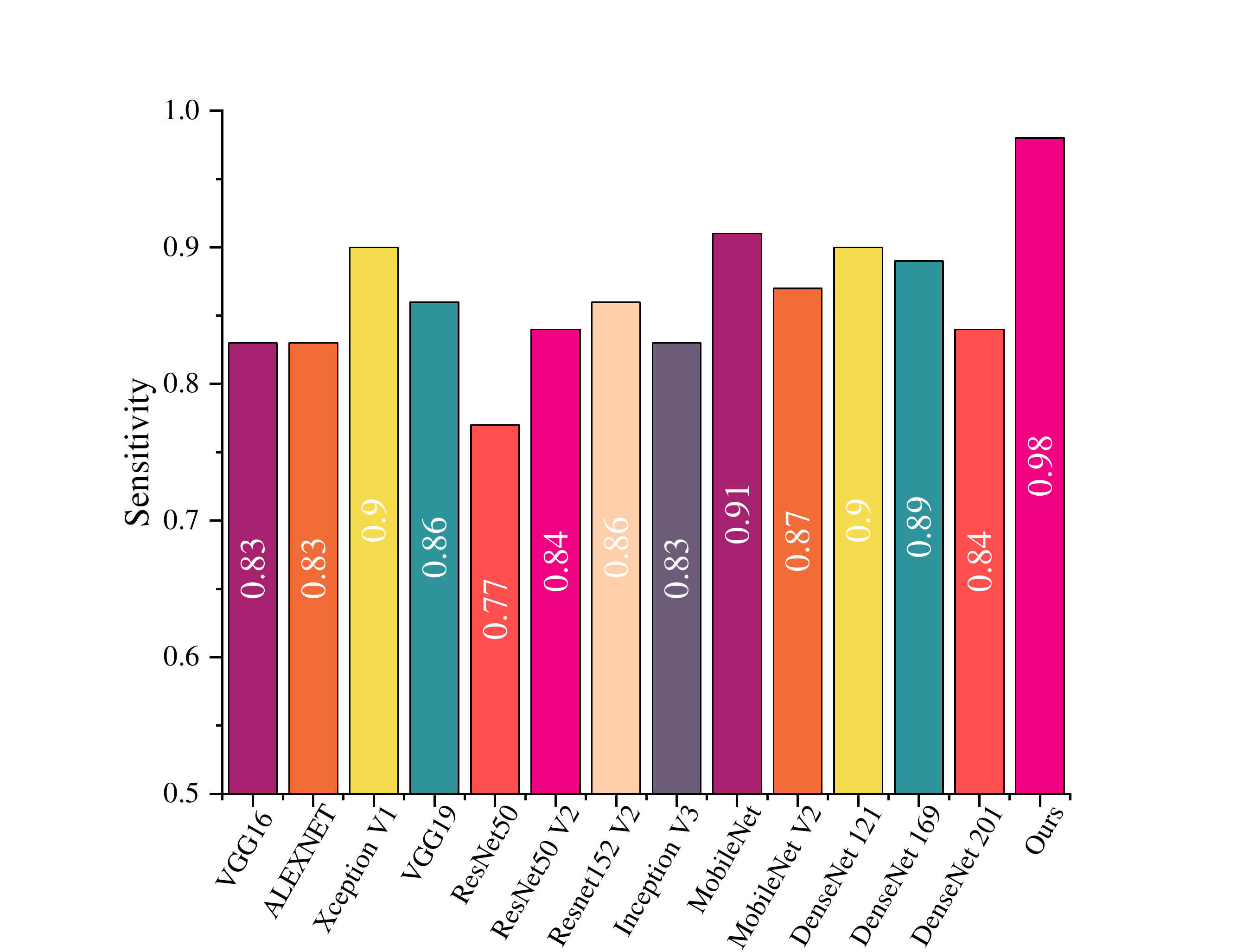}
	\caption{Sensitivity/Recall of the COVID-19 dataset over the decentralized network.}
	\label{fig:Senstivity-Specivity-Local-model}
\end{figure}

\begin{figure}[!tp]
	\centering
	\includegraphics[width=0.99\columnwidth]{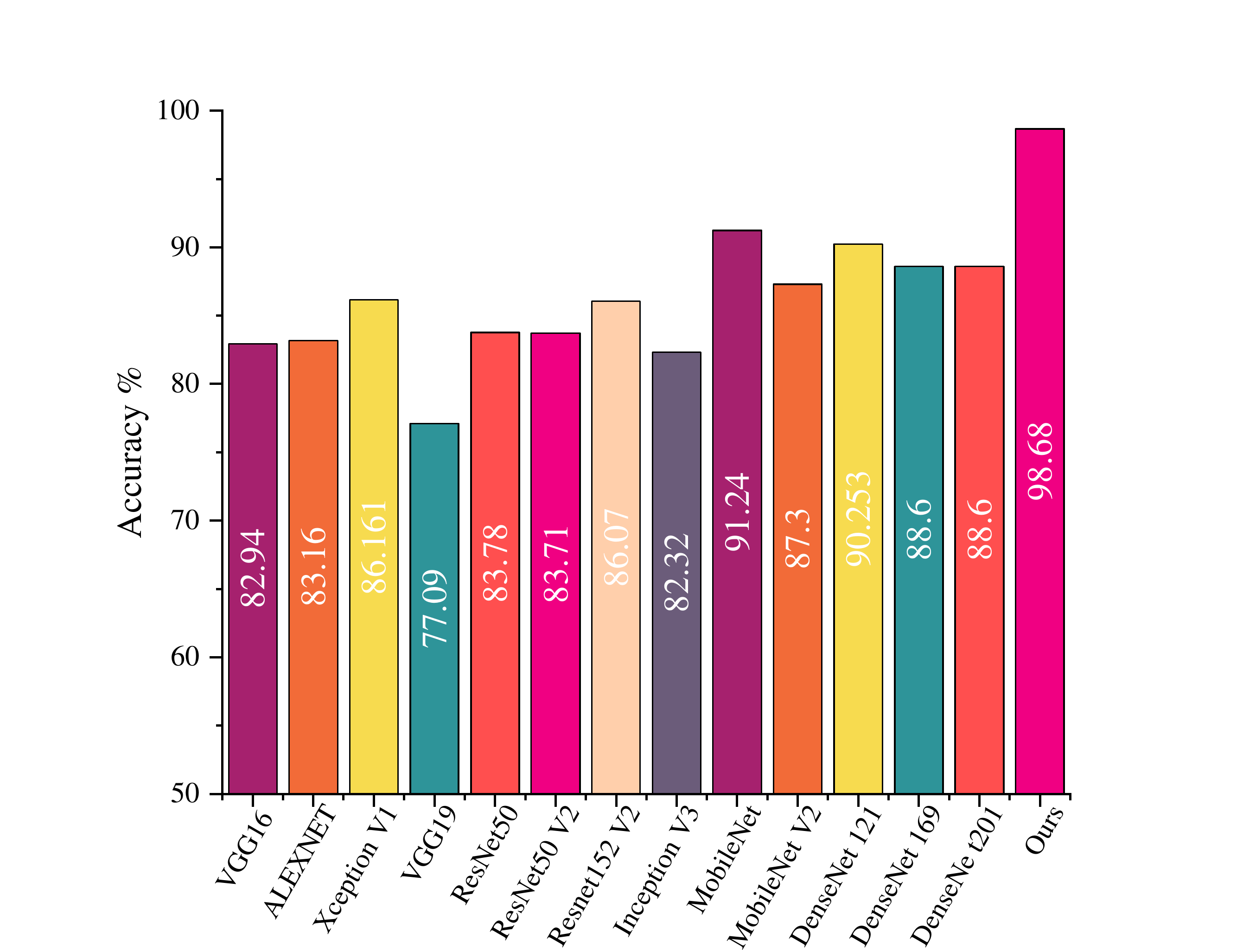}
	\caption{Accuracy of the COVID-19 Images.}
	\label{fig:Local-model-acc}
\end{figure}

\begin{figure*}[!tp]
	\centering
	\includegraphics[width=2.0\columnwidth]{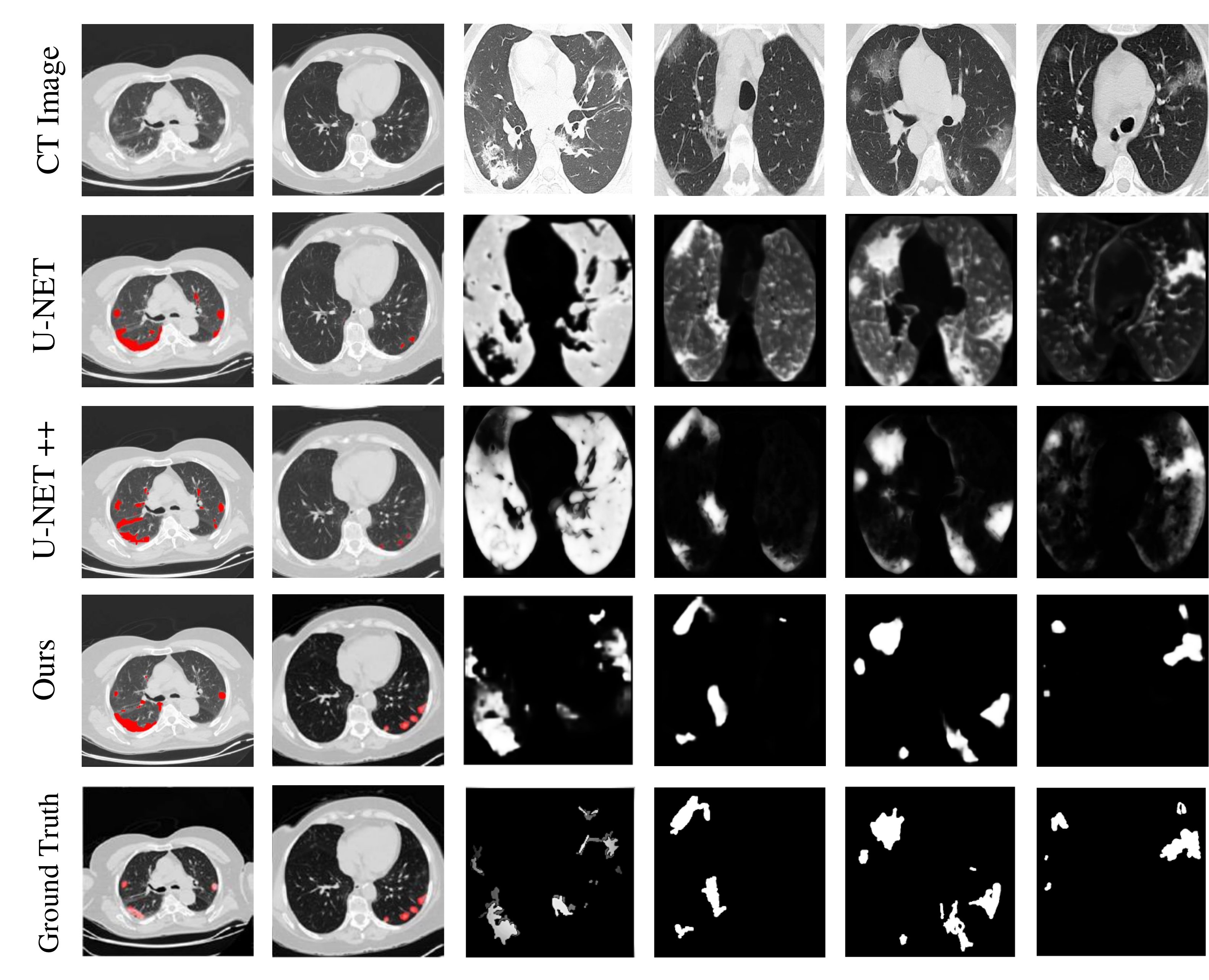}
	\caption{A visual comparison (segmentation) of our method with other studies. The first row (CT-Images) represents the original images taken from different datasets. The first two columns show the overlay segmentation result while the rest of the columns represent the masks.}
	\label{fig:seg}
\end{figure*}

We performed comprehensive experiments using different kinds of deep learning models i.e.,(VGG16, AlexNet, Inception V3, ResNet 50-152 layers, MobileNet, DenseNet). We used deep learning models and different layers for comparing the performance models on the COVID-19 dataset, which is shown in Table \ref{tab:True-Positive-local-model}. We evaluate the performance of the Capsule Network for the detection of COVID-19 lung CT image accuracy. Figure \ref{fig:Senstivity-Specivity-Local-model} shows the deep learning models; the Capsule Network achieves high sensitivity and less specificity, we achieved high detection performance through the Capsule Network. Figure \ref{fig:Local-model-acc} shows the Segcaps based Capsule Network achieved the best performance and provide the highest sensitivity and lowest specificity. These models were tested using three different test lists containing about 11,450 CT scan slices. The COVID-19 infection segmentation shown in Figure \ref{fig:seg}, indicates our method outperforms the baseline methods. The proposed techniques’ results are close to the ground truth. In contrast, U Net++s’ performance is near to our results.

\begin{table*}[t]
	\centering
	\caption{The performance of some famous deep learning networks. The bold values represent the best performance. It can be seen that the Capsule Network exhibited the highest sensitivity while ResNet has the best specificity.}
	
	\begin{tabular}{llllll}
		\midrule
		\toprule 
		
		\rule[-1ex]{0pt}{3.5ex}Feature extraction network & Learnable node & Pre-trained & Precision & Sensitivity / Recall &Specicivity\tabularnewline
		\hline
		\rule[-1ex]{0pt}{3.5ex}VGG16 \cite{DBLP:journals/corr/SimonyanZ14a} & MLP & Imagenet &0.8269	&0.8294&0.1561\tabularnewline
		\hline
		\rule[-1ex]{0pt}{3.5ex}AlexNet \cite{DBLP:conf/nips/KrizhevskySH12} & MLP & Scratch &\textbf{0.833}&	0.831&	0.191\tabularnewline
		\hline
		\rule[-1ex]{0pt}{3.5ex}Xception V1 \cite{DBLP:conf/cvpr/Chollet17} & MLP & Imagenet &0.830&	0.894&	0.110\tabularnewline
		\hline
		\rule[-1ex]{0pt}{3.5ex}VGG19 \cite{DBLP:journals/corr/SimonyanZ14a}& MLP & Imagenet &0.827&	0.8616&	0.128\tabularnewline
		\hline
		\rule[-1ex]{0pt}{3.5ex}ResNet50 \cite{DBLP:journals/corr/HeZRS15} & MLP & Imagenet &\textbf{0.833}	&	0.771&	\textbf{0.249}\tabularnewline
		\hline
		\rule[-1ex]{0pt}{3.5ex}ResNet50 V2 \cite{DBLP:conf/eccv/HeZRS16} & MLP & Imagenet &0.830	&0.837&	0.166\tabularnewline
		\hline
		\rule[-1ex]{0pt}{3.5ex}Resnet152 V2\cite{DBLP:conf/eccv/HeZRS16} & MLP & Imagenet &0.828	&	0.861&	0.134\tabularnewline
		\hline
		\rule[-1ex]{0pt}{3.5ex}Inception V3 \cite{DBLP:conf/cvpr/SzegedyVISW16} & MLP & Imagenet &	0.828&	0.833&	0.159
		\tabularnewline
		\hline
		\rule[-1ex]{0pt}{3.5ex}MobileNet \cite{DBLP:journals/corr/HowardZCKWWAA17} & MLP & Imagenet &0.830&	0.912&	0.089\tabularnewline
		\hline
		\rule[-1ex]{0pt}{3.5ex}MobileNet V2 \cite{SandlerHZZC18} & MLP & Imagenet &0.828	&	0.873&	0.118\tabularnewline
		\hline
		\rule[-1ex]{0pt}{3.5ex}DenseNet121 \cite{huang2017densely} & MLP & Imagenet &0.832&		0.903	&0.113\tabularnewline
		\hline
		\rule[-1ex]{0pt}{3.5ex}DenseNet169 \cite{huang2017densely} & MLP & Imagenet &0.831	&	0.886	&0.126\tabularnewline
		\hline
		\rule[-1ex]{0pt}{3.5ex}DenseNet201 \cite{huang2017densely} & MLP & Imagenet &0.829	&	0.844&	0.152\tabularnewline
		\hline
		\rule[-1ex]{0pt}{3.5ex}Ours (SegCaps) & Capsule Network & - &0.830 &\textbf{0.987}	&0.004\tabularnewline
		\hline
		\bottomrule
	\end{tabular}
	\label{tab:True-Positive-local-model}
\end{table*}

\subsection{Federated Learning Security analysis and Results}
\begin{figure}[!tp]
	\centering
	\includegraphics[width=0.99\columnwidth]{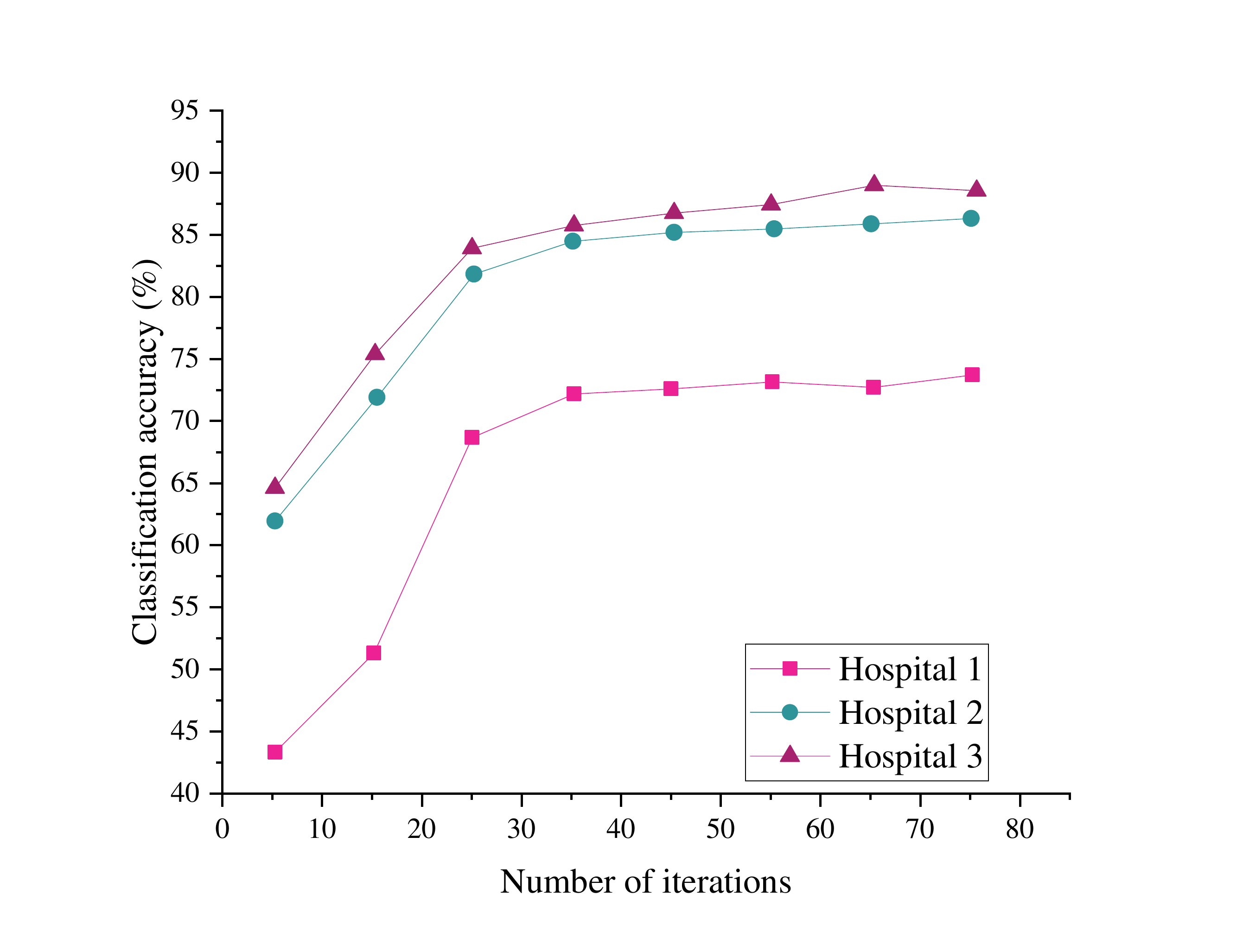}
	\caption{The accuracy of dataset COVID-19 for different providers.}
	\label{fig:Federated-ACC}
\end{figure}

\begin{figure}[!tp]
	\centering
	\includegraphics[width=0.99\columnwidth]{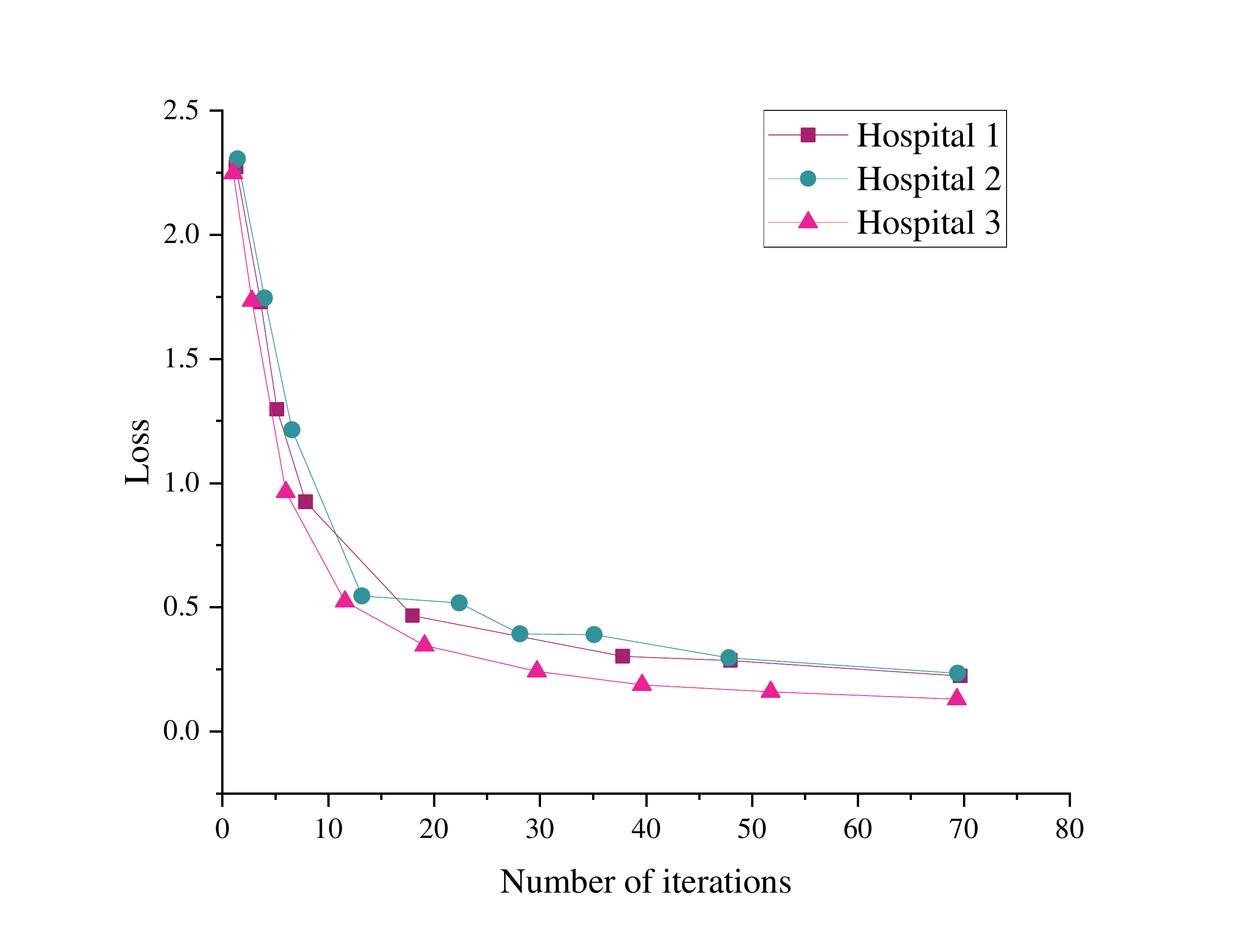}
	\caption{The Loss of dataset COVID-19 for different providers.}
	\label{fig:Federated-loss}
\end{figure}

\begin{figure}[!tp]
	\centering
	\includegraphics[width=0.99\columnwidth]{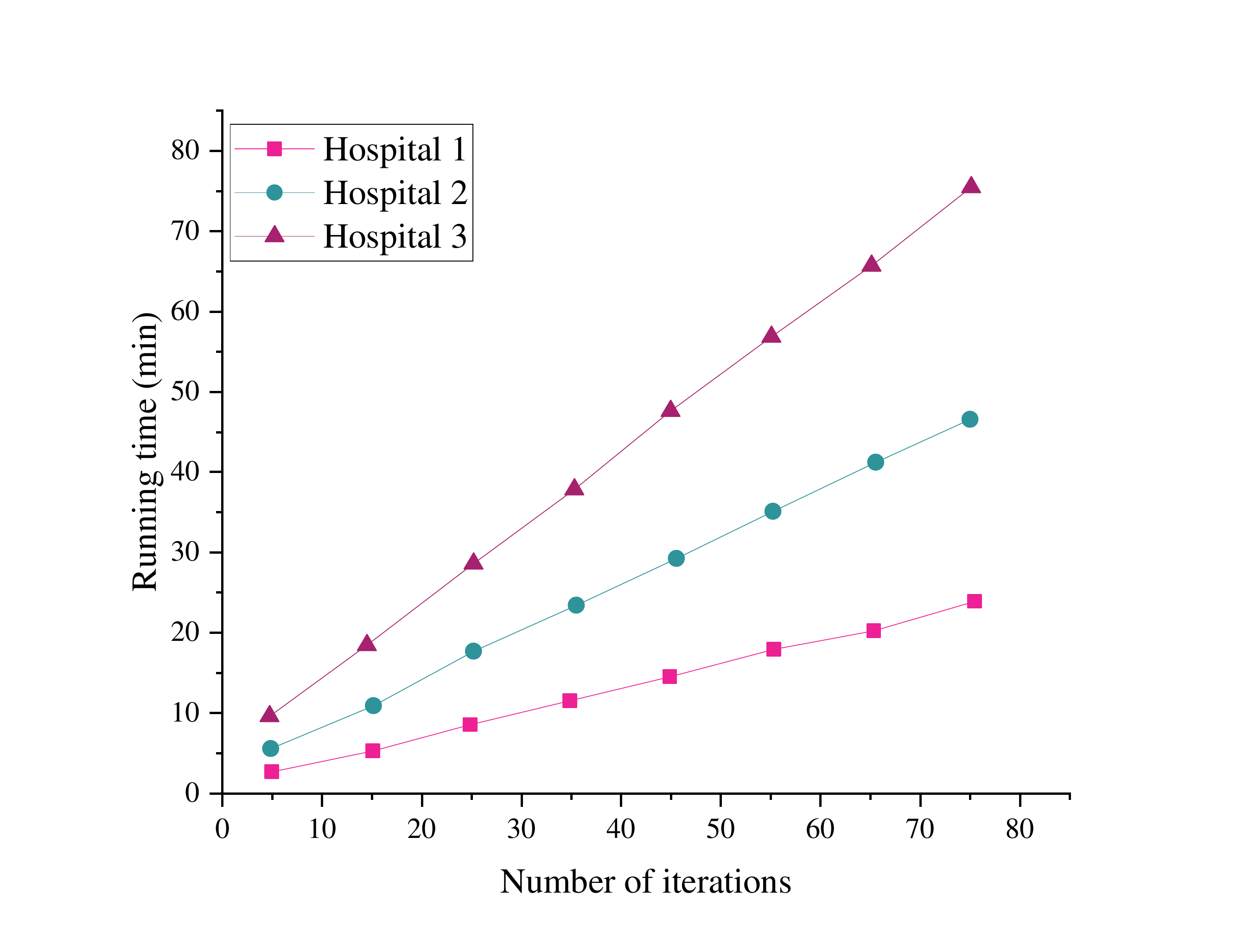}
	\caption{The time of dataset COVID-19 for different providers.}
	\label{fig:Federated-time}
\end{figure}

As  the dataset has been gathered from different sources and different hospitals having various kinds of machines. To measure the performance of federated learning, we distribute the datasets over three hospitals. In this model, multiple hospitals can share the data and learn from federated learning. The performance of our proposed model is distributed shown in Figure \ref{fig:Federated-ACC}, accuracy was changed when the hospitals or providers were increases. It is better to use more data providers for better results. Figure \ref{fig:Federated-loss} shows that model loss convergence. As we can see in Figure \ref{fig:Federated-ACC} the accuracy does not change smoothly because the samples from different hospitals are not the same. The accuracy depends on the number of patients or slices. The same is the process for the model loss. Also, it can be seen that the number of providers is increasing. The global model aggregate the local models, each local model normalizes the data before training a local model. The number of hospitals affects the performance of the collaborative model. Additionally, the run time is shown in Figure \ref{fig:Federated-time}. It varies for the datasets and number of iteration in different sub-datasets.

We compare the federated learning with the local model as shown in Figure \ref{fig:Local-model-acc}. The local model is trained on the whole dataset and the federated learning model learns from the local models. Figure \ref{fig:Federated-ACC} and \ref{fig:Federated-loss} indicates that performance increases significantly when data providers are increasing. However, federated learning does not affect the accuracy but it achieves privacy while sharing the data.

\begin{itemize}
	\item \textbf{Differences-Privacy:} Figure \ref{fig:Private-Federated-Learning} describes the differences in privacy analysis, where a principled approach that enables organizations to learn from most data while ensuring that these results do not allow data to be distinguished or re-identified by any individual. On the other hand, Equation \ref{eq:3} obtains the value in the data to ensure strong data security.
	\item \textbf{Trust: }The decentralized trust mechanism of the blockchain allows everything to run automatically through a preset program, which improves data security. Relying on a strict set of algorithms, the decentralized blockchain technology can ensure that the data is true, accurate, transparent, traceable, and cannot be tampered with.
	\item \textbf{ Data security:} Data providers have the authority to control their data. Actual data is uploaded with the signature of the owner in the blockchain database. The owner has the right to control and change the policy of the data using the smart contract. The blockchain uses cryptographic algorithms that enable the security of the data.

\end{itemize}
\subsection{Comparison with other methods}
A lot of previous studies have been carried out for detecting the COVID-19 such as \cite{Kuzdeuov2020,Wang2020, Zhou2020, Wang2020a, Sun2020}, these techniques do not consider data sharing to train the better prediction model. However, some techniques used GAN and data augmentation for generating fake images. The performance of such methods is not reliable in the case of medical images. Due to the small number of data patients \cite{inciardi2020cardiac} the data analytic is difficult. Our proposed model collects a huge amount of real-time data to build a better prediction model. 
Firstly, we compare with the state of art studies and compare them with the deep learning models shown in Figure \ref{tab:compare}. Moreover, we compare federated learning with the state-of-art deep learning models such as VGG, RESNET, ImageNet, MobileNet, Desnet, Capsule Network. The results show the accuracy is similar to train the local model with the whole dataset or divide data into different hospitals and combine the model weights using blockchain based federated learning.

Finally, we compare our work with blockchain-based data sharing techniques. \cite{kumar2020integration} proposed a deep learning and blockchain-based technique to share the medical images, but the main weakness of the model is that it is not based on federated learning and do not aggregates the neural network weights over the blockchain. Moreover, \cite{kong2019federated,lu2020blockchain} design a framework based on federated learning but they only consider share vehicle data. Our proposed framework trains the global model to collect data from different hospitals and train a collaborative global model.


\begin{table*}
	
	\caption{A comparison with the state-of-the-art studies related to COVID-19 patients’ detection}
	\label{tab:compare}
	\begin{tabular}{>{\raggedright}p{0.25\columnwidth}>{\raggedright}p{0.3\columnwidth}>{\raggedright}p{0.35\columnwidth}>{\raggedright}p{0.3\columnwidth}>{\raggedright}p{0.3\columnwidth}>{\raggedright}p{0.1\columnwidth}>{\raggedright}p{0.12\columnwidth}}
		\midrule
		\toprule 
		Study & Backbone method & Tasks & Number of cases & Performance & Sharing & Privacy protection\tabularnewline
		\midrule
		
		Chen et al. \cite{chen2020deep} & 2D Unet++ & COVID-19, viral, baterial and Pneu. classification & 106 cases & 95.2 \%(Accuracy)
		
		100\% (Senstivity)
		
		93.6\%(Specificity) & No & No\tabularnewline
		\midrule 
		Shi et al. \cite{shi2020large} & Random Forest & COVID-19, viral/ baterial Pneu. and normal classification & 2685 cases & 87.9 \%(Accuracy)
		
		90.7\% (Senstivity)
		
		83.3\%(Specificity) & No & No\tabularnewline
		\midrule 
		Zheng et al. \cite{zheng2020deep} & 2D Unet and 2D CNN & COVID-19, viral, baterial and Pneu. classification & 499 Training / 132 Validation & 90.7\% (Senstivity)
		
		91.1\%(Specificity) & No & No\tabularnewline
		\midrule 
		Li et al. \cite{li2020using}& 2D-Resnet 50 & COVID-19, viral/ baterial Pneu. and normal classification & 3920 Training/436 Testing & 90.0\% (Senstivity)
		
		96.0\%(Specificity) & No & No\tabularnewline
		\midrule 
		Jin et al. \cite{jin2020ai} & 2D Unet++ and 2D-CNN & COVID-19, viral, baterial and Pneu. classification & 1136 Training / 282 Testing & 97.4\% (Senstivity)
		
		92.2\%(Specificity) & No & No\tabularnewline
		\midrule 
		Song et al \cite{song2020deep}. & 2D-Resnet 50 & COVID-19, baterial Pneu. and normal classification & 164 Training/ 27 Validation/ 83 Testing & 86.0\% (Accuracy) & No & No\tabularnewline
		\midrule 
		Xu et al. \cite{xu2020deep} & 2D-CNN & Normal, Influenza-A and viral/bacterial Pneu. classification & 528 Training / 90 Testing & 86.7\% (Accuracy) & No & No\tabularnewline
		\midrule 
		Jin et al.\cite{jin2020development} & 2D-CNN & COVID-19, viral, baterial and Pneu. classification & 312 Training/ 104 Validation/ 1255 Testing & 94.1\% (Senstivity)
		
		95.5\%(Specificity) & No & No\tabularnewline
		\midrule 
		Wang et al. \cite{wang2020deep} & 2D-CNN & COVID-19 and viral Pneu. classification & 250 cases & 82.9\% (Accuracy) & No & No\tabularnewline
		\midrule 
		Wang et al. \cite{wang2020prior} & 3D-ResNet + attention & COVID-19, viral Pneu. and normal classification & 3997 5-fold validation / 60 validation/ 600 testing & 93.3 \%(Accuracy)
		
		87.6\% (Senstivity)
		
		95.5\%(Specificity) & No & No\tabularnewline
		\midrule 
		Ours & Federated Blockchain and Capsule Network & COVID-19, viral Pneu. and normal classification & 182 Training/ 45 Testing (patients per hospital) & 98.68\%(Accuracy)
		
		98\% (Senstivity)
		& Yes & Yes\tabularnewline
		\midrule
		\midrule 
		& & & & & & \tabularnewline
	\end{tabular}

\end{table*}

\section{Related Work\label{sec:Related-Work}}

\subsection{AI in COVID-19 }

Artificial Intelligence (AI) based techniques have played an essential role in the domain of medical image processing, computer-aided diagnosis, image interpretation, image fusion, image registration, image segmentation, image-guided therapy, image retrieval, and analysis techniques. Artificial Intelligence aids in extracting information from the images and represent information effectively and efficiently. Artificial Intelligence facilitates and assists doctors and other medical practitioners to diagnose various diseases while eliminating human error and increasing the speed and accuracy of detection. These techniques enhance the abilities of doctors and researchers to understand how to analyze the generic variations which cause the disease in the first place. Deep learning is the core technology of the rising artificial intelligence and has reported significantly diagnostic accuracy in medical imaging for automatic detection of lung diseases \cite{Ardila2019,Suzuki2017,Coudray2018} . Deep learning surpassed human- performance on the ImageNet image classification task, with one million images for training in 2015 \cite{He2015}, which showed dermatologist-level performance on classifying skin lesions in 2017 \cite{Esteva2017} and obtained remarkable results for lung cancer screening in 2019 \cite{Ardila2019}.
Pneumonia can be diagnosed using Computed Tomography (CT) scans of the chest of the subject. Artificial Intelligence (AI) based automated CT image analysis tools for the detection, quantification, and monitoring of coronavirus and to distinguish patients with coronavirus from disease-free have been developed \cite{Negassi2020}. In a study by Fei et al., they developed a deep learning-based system for automatic segmentation of all lung and infection sites using chest CT \cite{Gulshan2016}. Xiaowei et al. aimed to establish an early screening model to distinguish COVID-19 pneumonia and Influenza-A viral pneumonia from healthy cases using pulmonary CT images and deep learning techniques \cite{Ai2020}. Shuai et al., their study was based on the COVID-19 radiographic changes from CT images. They developed a deep learning method that can extract the graphical features of COVID-19 to provide a clinical diagnosis before pathogenic testing and thus save critical time for the disease diagnosis. Recently, C. Zheng et al. \cite{Zheng2020} developed a deep learning-based model for automatic COVID-19 detector using 3D CT volumes.

\subsection{Federated Learning}

Federated learning was proposed by McMahan et al. \cite{BrendanMcMahan2017} to learn from the shared model while protecting the privacy of data. In this context, federated learning is used to secure data and aggregates the parameters for the multiple organizations\cite{lu2020privacy, yan2020federated, ye2020federated, mowla2019federated, brik2020federated}. The hospitals can share the dataset during training and information about their dataset is revealed through analyzing the distributed model \cite{McMahan2017a,Lu2020,Lu2020a,Yin2020,Lim2020,Lin2020,Yang2020,Kim2019a,Yang2019}. This decentralized approach to train models preserves privacy and security. A lot of research has been done in federated learning for transferring the matrices of weights of deep neural networks. The previous studies do not consider to share the medical data without compromising the privacy of organizations\cite{Li2019a,Li2019}. In this article, we simulate our model to collect the data from different sources using federated learning combined with blockchain technology while sharing data without privacy leakage.

\section{Conclusion}

This paper proposed a framework that can utilize up-to-date data to improve the recognition of computed tomography (CT) images and share the data among hospitals while preserving privacy. The data normalization technique deals with the heterogeneity of data. Further, Capsule Network based segmentation and classification is used to detect COVID-19 patients along with a method that can collaboratively train a global model using blockchain technology with federated learning. Also, we collected real-life COVID-19 patients’ data and made it publically available to the research community. Extensive experiments were performed on various deep learning models for training and testing the datasets. The Capsule Network achieved the highest accuracy. The proposed model is smart as it can learn from the shared sources or data among various hospitals. Conclusively, the proposed model can help in detecting COVID-19 patients using lung screening as hospitals share their private data to train a global and better model.

	\section*{Declaration of Competing Interest}
	The au­thors re­port no de­c­la­ra­tions of in­ter­est.
	\section*{Acknowledgement}
	This work was sup­ported by China Post­doc­toral scine Sci­ence Foun­da­tion and De­part­ment of Sci­ence and Tech­nol­ogy of Sichuan Province , Pro­ject Num­ber: Y03019023601016201, H04W200533.
	Authors’ contributions

	\bibliographystyle{IEEEtran}
	\bibliography{ref}
	
	\begin{IEEEbiography}[{\includegraphics[width=1in,height=1.25in,clip,keepaspectratio]{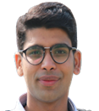}}]{Rajesh Kumar} was born in Sindh Province of Pakistan in November 1991. He received his B.S. and M.S. degree in computer science from University of Sindh, Jamshoro, Pakistan. He received his Ph.D. in computer science and engineering from the University of Electronic Science and Technology of China (UESTC). He has currently pursing Post Doctor in Information Security at the University of Electronic Science and Technology of China. His research interests include machine learning, deep leaning, malware detection, Internet of Things (IoT) and blockchain technology. In addition, he has published more than 20 articles in various International journals and conference proceedings.
	\end{IEEEbiography}
	
	\vspace{-1.0 cm}
	\begin{IEEEbiography}[{\includegraphics[width=1in,height=1.25in,clip,keepaspectratio]{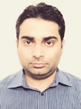}}]{Abdullah Aman Khan} 		Khan received his master’s degree in the field of Computer Engineering from National University of Science and Technology (NUST), Punjab, Pakistan in 2014. He is currently perusing PhD degree in the field of Computer Science and Technology form the school of computer science and engineering, University of Electronic Science and Technology, Sichuan, Chengdu, P.R. China. His main area of research includes electronics design, machine vision and intelligent systems.
		
\end{IEEEbiography}
	\vspace{-1.0 cm}
	\begin{IEEEbiography}[{\includegraphics[width=1in,height=1.25in,clip,keepaspectratio]{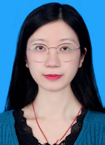}}]{Simin Zhang}	received her Master’s degree in Sichuan University, China. Currently, she is pursuing her Ph.D. in Medical imaging and nuclear medicine from West China Hospital of Sichuan University. Her research interests include machine learning and multi-model MRI applied to glioma.
		
\end{IEEEbiography}
	\vspace{-1.0 cm}
	
	\begin{IEEEbiography}[{\includegraphics[width=1in,height=1.25in,clip,keepaspectratio]{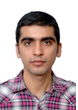}}]{Jay Kumar}	 received his Master’s degree in Computer Science from Quaid-i-Azam University, Pakistan. Currently, he is pursuing his Ph.D. in computer science and engineering from the University of Electronic Science and Technology of China (UESTC). His research interests include text classification and subspace clustering in stream mining applied to various domains.
	\end{IEEEbiography}
	\vspace{-1.0 cm}

	\begin{IEEEbiography}[{{\includegraphics[clip,width=1in,height=1.25in]{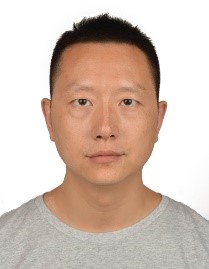}}}]{Professor Ting Yang }		He obtained Bachelor Degree of communication engineering from UESTC; He obtained Master Degree from School of Computer Science and Engineering		of UESTC, His Ph.D. in electronics engineering from the University		of Electronic Science and Technology of China (UESTC). His research interests include computer network architecture, software defined networking, and Internet of Things (IoT).
\end{IEEEbiography}
	
\vspace{-1.0 cm}	\begin{IEEEbiography}[{{\includegraphics[clip,width=1in,height=1.25in]{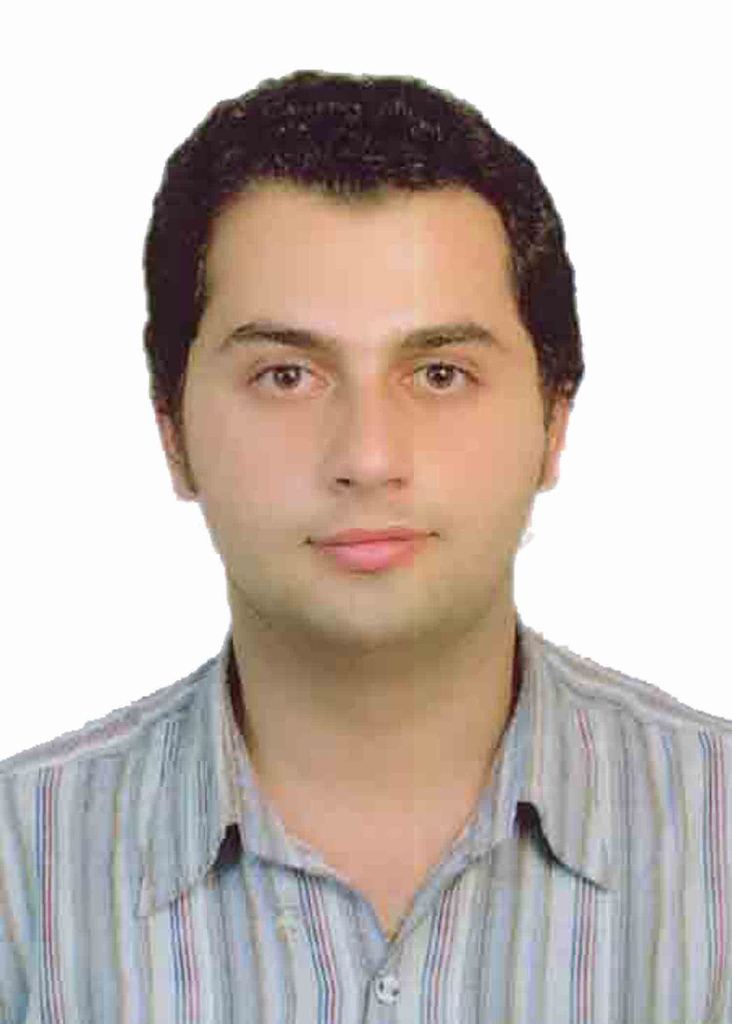}}}]{Noorbakhsh Amiri Golilarz } received the M.Sc. degree from Eastern Mediterranean University, Cyprus, in 2017. He is currently pursuing the Ph.D. degree with the School of Computer Science and Engineering, University of Electronic Science and Technology of China (UESTC). His main research interests include image processing, satellite and hyper-spectral image de-noising, biomedical, signal processing, optimization algorithms, control, systems, pattern recognition, neural networks, and deep learning.
	\end{IEEEbiography}
	
	\begin{IEEEbiography}[{{\includegraphics[clip,width=1in,height=1.25in]{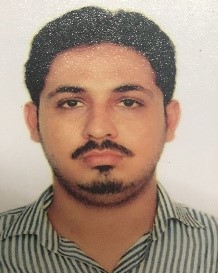}}}]{Zakria}		received the M.S. degree in Computer Science and Information from		N.E.D University in 2017. He is currently pursuing the Ph.D. degree	with the School of Information and Software Engineering, University of Electronic Science and Technology of China. He has a vast academic, technical, and professional experience in Pakistan. His research interests include artificial intelligence, computer vision particularly vehicle re-identification 
	\end{IEEEbiography}
	


	\vspace{-1.0 cm}
	\begin{IEEEbiography}[{\includegraphics[width=1in,height=1.25in,clip,keepaspectratio]{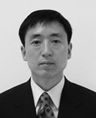}}]{Professor Wenyong Wang} 	was born in 1967. Currently, He is working as a professor of Computer Science at University of Electronic Science and Technology of China (UESTC). He holds a B.E. in Computer Science from Beijing University of Aeronautics and Astronautics, Beijing, China, an M.E. in Computer Science and a Ph.D. in Communications Engineering from UESTC. His research interests include computer network architecture, software defined networking, and Internet of Things (IoT).
	\end{IEEEbiography}


\end{document}